\documentclass[12pt]{article}

\usepackage{amsmath,amsfonts,enumerate,array}
\usepackage{graphicx}
\usepackage[usenames]{color}
\usepackage[english]{babel}
\usepackage{fancybox}
\usepackage{chngpage} 

\usepackage{tikz-cd}

\newcommand{\captionfonts}{\small}

\makeatletter  
\long\def\@makecaption#1#2{%
  \vskip\abovecaptionskip
  \sbox\@tempboxa{{\captionfonts #1: #2}}%
 \ifdim \wd\@tempboxa >\hsize
    {\captionfonts #1: #2\par}
  \else
    \hbox to\hsize{\hfil\box\@tempboxa\hfil}%
  \fi
  \vskip\belowcaptionskip}
\makeatother   

\usepackage{mciteplus}

\usepackage[colorlinks=true,      linkcolor=blue,      urlcolor=blue,      filecolor=blue,      citecolor=blue,
      pdfstartview=FitH,     pdfpagemode=UseNone,      bookmarksopen=true]{hyperref}  
      
\usepackage[all]{hypcap}     

\usepackage{cite}

\begin{document}

\numberwithin{equation}{section}


\mathchardef\mhyphen="2D


\newcommand{\be}{\begin{equation}} 
\newcommand{\ee}{\end{equation}} 
\newcommand{\bea}{\begin{eqnarray}\displaystyle}
\newcommand{\eea}{\end{eqnarray}}
\newcommand{\bt}{\begin{tabular}}
\newcommand{\et}{\end{tabular}}
\newcommand{\bs}{\begin{split}}
\newcommand{\es}{\end{split}}

\newcommand{\I}{\text{I}}
\newcommand{\II}{\text{II}}

\renewcommand{\a}{\alpha}	
\renewcommand{\b}{\beta}
\newcommand{\g}{\gamma}		
\newcommand{\G}{\Gamma}
\renewcommand{\d}{\delta}
\newcommand{\D}{\Delta}
\renewcommand{\c}{\chi}			
\newcommand{\C}{\Chi}
\newcommand{\p}{\psi}			
\renewcommand{\P}{\Psi}
\newcommand{\s}{\sigma}		
\renewcommand{\S}{\Sigma}
\renewcommand{\t}{\tau}		
\newcommand{\e}{\epsilon}
\newcommand{\n}{\nu}
\newcommand{\m}{\mu}
\renewcommand{\r}{\rho}
\renewcommand{\l}{\lambda}

\newcommand{\nn}{\nonumber\\} 		
\newcommand{\newotimes}{}  				
\newcommand{\diff}{\,\text{d}}		
\newcommand{\h}{{1\over2}}				
\newcommand{\Gf}[1]{\G \Big{(} #1 \Big{)}}	
\newcommand{\floor}[1]{\left\lfloor #1 \right\rfloor}
\newcommand{\ceil}[1]{\left\lceil #1 \right\rceil}

\def\cA{{\cal A}} \def\cB{{\cal B}} \def\cC{{\cal C}}
\def\cD{{\cal D}} \def\cE{{\cal E}} \def\cF{{\cal F}}
\def\cG{{\cal G}} \def\cH{{\cal H}} \def\cI{{\cal I}}
\def\cJ{{\cal J}} \def\cK{{\cal K}} \def\cL{{\cal L}}
\def\cM{{\cal M}} \def\cN{{\cal N}} \def\cO{{\cal O}}
\def\cP{{\cal P}} \def\cQ{{\cal Q}} \def\cR{{\cal R}}
\def\cS{{\cal S}} \def\cT{{\cal T}} \def\cU{{\cal U}}
\def\cV{{\cal V}} \def\cW{{\cal W}} \def\cX{{\cal X}}
\def\cY{{\cal Y}} \def\cZ{{\cal Z}}

\def\mC{\mathbb{C}} \def\mP{\mathbb{P}}  
\def\mR{\mathbb{R}} \def\mZ{\mathbb{Z}} 
\def\mT{\mathbb{T}} \def\mN{\mathbb{N}}
\def\mH{\mathbb{H}} \def\mX{\mathbb{X}}
\def\CP{\mathbb{CP}}
\def\RP{\mathbb{RP}}
\def\Z{\mathbb{Z}}
\def\N{\mathbb{N}}
\def\H{\mathbb{H}}

\def\b{\bigskip}

\begin{flushright}
\end{flushright}
\vspace{20mm}
\begin{center}
{\LARGE Bootstrapping the effect of the twist operator\\\vspace{2mm} in symmetric orbifold CFTs}
\\
\vspace{18mm}
\textbf{Bin} \textbf{Guo}{\footnote{bin.guo@ipht.fr}}~\textbf{and} ~ \textbf{Shaun}~\textbf{D.}~\textbf{Hampton}{\footnote{shaun.hampton@ipht.fr}}
\\
\vspace{10mm}

${}$Institut de Physique Th\'eorique,\\
	Universit\'e Paris-Saclay,
	CNRS, CEA, \\ 	Orme des Merisiers,\\ Gif-sur-Yvette, 91191 CEDEX, France  \\

\vspace{8mm}
\end{center}

\vspace{4mm}

\thispagestyle{empty}

\begin{abstract}
We study the 2D symmetric orbifold CFT of two copies of free bosons. The twist operator can join the two separated copies in the untwisted sector into a joined copy in the twisted sector. Starting with a state with any number of quanta in the untwisted sector, the state in the twisted sector obtained by the action of the twist operator can be computed by using the covering map method. We develop a new method to compute the effect of a twist operator by using the Bogoliubov ansatz and conformal symmetry. This may lead to more efficient tools to compute correlation functions involving twist operators.

\vspace{3mm}

\end{abstract}
\newpage

\setcounter{page}{1}

\numberwithin{equation}{section} 

\tableofcontents

\newpage

\section{Introduction}

Symmetric orbifold CFTs are widely used in AdS$_3$/CFT$_2$ as boundary theories to understand the bulk physics \cite{Maldacena:1997re,Strominger:1996sh,Maldacena:1999bp,Seiberg:1999xz,Dijkgraaf:1998gf,Larsen:1999uk,Jevicki:1998bm,deBoer:1998kjm}. Consider a seed CFT with target space $M$. The symmetric orbifold CFT has the target space
\be
M^N/S_N
\ee
where $N$ is the number of copies of the seed CFT and $S_N$ is the permutation group of $N$ elements.

Due to the $S_N$ orbifold, there exist twist operators. If we circle the insertion of a twist operator, different copies of the seed CFT permute into each other. The traditional way to compute correlation functions of twist operators is to map the 2-d base space to the covering space \cite{Lunin:2000yv,Lunin:2001pw,Pakman:2009zz,Pakman:2009ab}. On this covering space, the ramification caused by the twist operator in the base space is resolved, such that the target space becomes a single copy of $M$. 
The correlation functions on the base space can be computed from a combination of the correlation functions in the covering space and a Liouville factor which takes into account the covering map.

\begin{figure}
\centering
        \includegraphics[width=9.5cm]{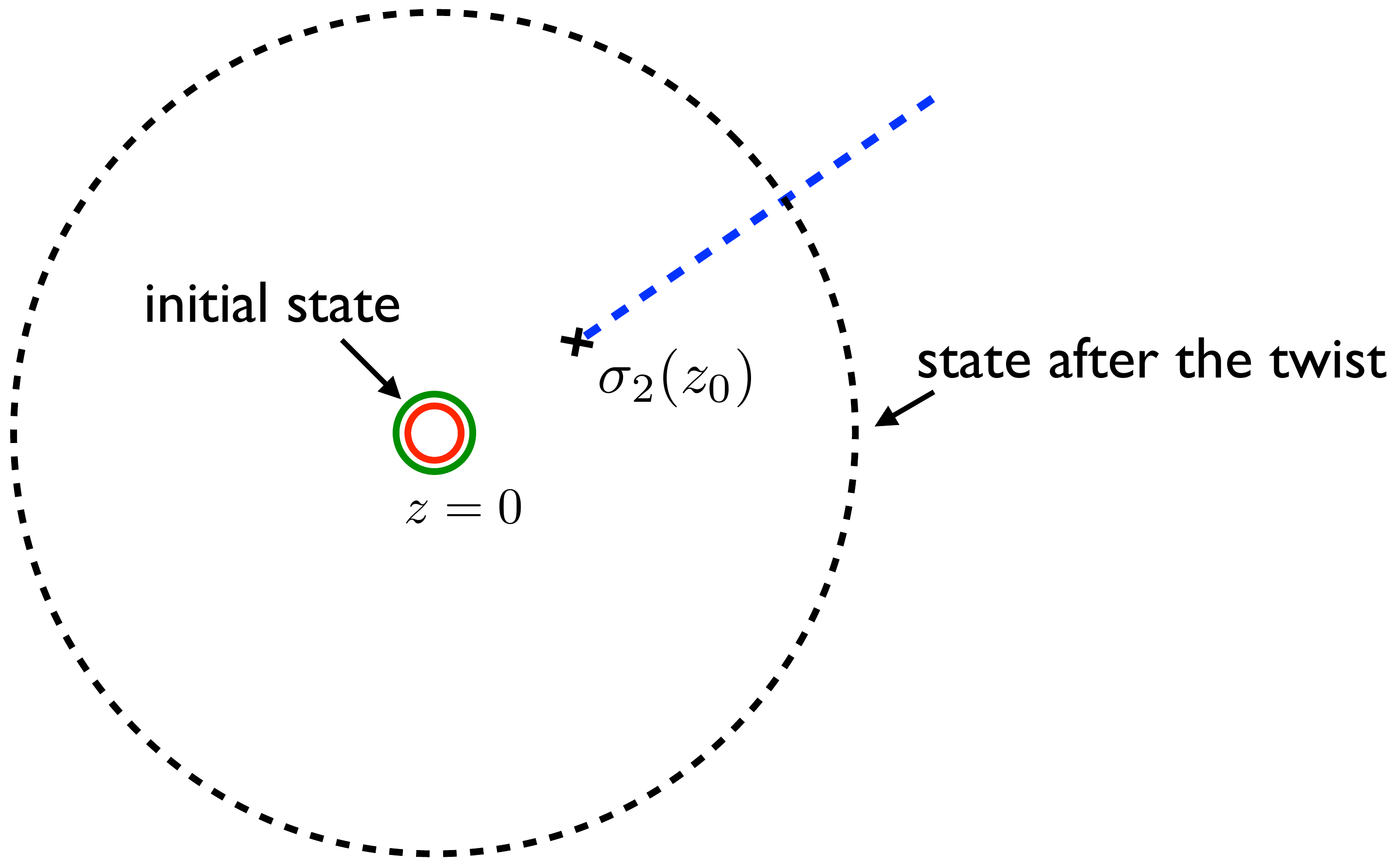}
\caption{The effect of a twist operator. The red and green circles represent the states living on two separate copies located before the twist operator. The twist operator produces a branch cut from $z_0$ to infinity. The dashed circle represents the state living on the joined copy with two sheets.}
\label{fig}
\end{figure}

The effect of a twist operator can be studied by using the covering map \cite{Avery:2010er,Avery:2010hs,Avery:2010qw,Carson:2014yxa,Carson:2014ena}.
To be specific, take two copies of the seed CFT of a free boson, such that $M=R$. Suppose around $z=0$ that the state is in the untwisted sector as shown in fig \ref{fig}. Around this point, there are two separate copies of the seed CFT. Let us put a twist operator $\sigma_2$ at $z_0$, which is the unique twist operator for two copies. This twist operator generates a branch cut from $z_0$ to infinity. The two separate copies join into a single doubly wound copy for $|z|>|z_0|$. If an initial state is given at $z=0$ for the two separate copies, what is the state after the twist operator, e.g. the state at infinity? The result of this question tells us all the three-point functions involving a twist operator $\sigma_2$. 

It has been observed that the effect of a twist operator is in the form of a Bogoliubov transformation between the modes before and after the twist operator. From the covering space method point of view, the map from the base space to the covering space leads to this linear transformation of modes. The effect of a twist operator is encoded in this linear transformation. This can be derived from the covering map or by matching the modes just before and after the twist \cite{Carson:2014xwa,Carson:2017byr}. In the covering map method, it seems that the details of the map are necessary. In the latter method, the covering map is not needed but in practice not all effects can be obtained since it requires one to invert an infinite-dimensional matrix.

In this paper, we will develop a method that does not involve the covering map and can obtain the effects completely. To do that, we will use the `weak' Bogoliubov ansatz\footnote{ `Weak' means that we don't require some relations among the coefficients in the original Bogoliubov ansatz. For more details, see section \ref{sec B}. These relations come out naturally. As a result, this method does not involve the inversion of infinite-dimensional matrices.}
and conformal symmetry
\be
\text{Weak Bogoliubov ansatz} ~+~ \text{Conformal symmetry}~ \Rightarrow ~\text{Effect of a twist operator}
\ee
For earlier work in this direction see \cite{Burrington:2014yia} which incorporates the use of the covering map and conformal symmetry. 
There have also been works to compute correlation functions of twist operators using conformal symmetry, see e.g. \cite{Dixon:1986qv,Arutyunov:1997gt,Arutyunov:1997gi,Jevicki:1998bm}. For recent work, see \cite{Dei:2019iym}.

The plan of the paper is as follows: In section \ref{sec mode} we outline the orbifold CFT of one boson. In section \ref{sec ansatz} we describe the effect of the twist operator. In section \ref{sec B} we discuss the Bogoliubov transformation. In section \ref{bootstrap} we bootstrap the effect of the twist operator using the weak Bogoliubov ansatz and conformal symmetry. In section \ref{discussion} we discuss our results and future work.

\section{Orbifold CFT of one boson}\label{sec mode}

Symmetric orbifold CFTs are obtained by orbifolding $N$ copies of a seed CFT by the permutation group $S_{N}$, which results in the target space
\be
M^{N}/S_N
\ee
where $M$ is the target space of the seed CFT.
In this paper, we will consider the simplest case where the seed CFT is a free boson with target space $M=R$.
The base space is the complex $z$ plane.

The $N$ copies of the free boson are labeled by $X^{(i)}$ with $i=1,\dots,N$. In the untwisted sector, the fields have the boundary condition
\bea\label{untwist b}
X^{(i)}\to X^{(i)},~~~~z\to z e^{2\pi i}
\eea
There also exist twist sectors where the $N$ copies can join into many linked copies in all possible ways. For example, a $k$-wound linked copy has the boundary condition
\bea
X^{(1)}\to X^{(2)}\to\ldots\to X^{(k)}\to X^{(1)},~~~~z\to z e^{2\pi i}
\eea
It is convenient to define a single field $X$ living on the $k$-wound copy. On the $i$-th segment of the $k$-wound copy, the field $X$ equals to $X^{(i)}$, such that the field $X$ has the boundary condition
\be
X \to X,~~~~z\to z e^{2\pi k i}
\ee
Notice that the field $X$ is multi-valued in the base space and should be thought of as a single-valued field living on a Riemann surface with $k$ sheets.

In radial quantization, the modes of the holomorphic part in the untwisted sector are defined as
\be
\a^{(i)}_{n} = {1\over2\pi}\oint_{C_0}dz \, z^{n} \partial X^{(i)} (z)
\ee
where $C_0$ is a contour centered around $z=0$. 
The $n$ is an integer as required by the boundary condition (\ref{untwist b}).
The commutation relation is
\be\label{com untwist}
[\a^{(i)}_{m},\a^{(j)}_{n}]=m\delta^{ij}\d_{m+n,0}
\ee
We also define
\be
\a^{(i)\dagger}_n = \a^{(i)}_{-n}
\ee
The vacuum $|0\rangle^{(i)}$ of copy $i$ is defined by the condition
\be\label{v untwist}
\a^{(i)}_{n}|0\rangle^{(i)}=0, ~~~~n\geq 0
\ee
The Virasoro generators can be expanded in terms of a sum over bilinears of the modes
\be\label{L untwist}
L_{m} = {1\over2}\sum_{i}\sum_n\a^{(i)}_{n}\a^{(i)}_{m-n}
\ee
with implicit normal-ordering.
Using the commutation relation (\ref{com untwist}), we have
\be\label{4-9}
[L_m,\a^{(i)}_n]=-n \a^{(i)}_{m+n}
\ee

For the field $X$ living on the $k$-wound copy, the modes are defined as
\be
\a_{n} = {1\over2\pi}\oint_{C_0^{(2\pi k)}}dz \, z^{n} \partial X (z)
\ee
where the contour of the integral $C_0^{(2\pi k)}$ is again centered around $z=0$ but now from angle $0$ to $2\pi k$. The boundary condition for the field $X$ requires that $n=m/k$ where $m$ is an integer. The commutation relation is given by 
\be\label{comm twist}
[\a_{m},\a_{n}]=k m\d_{m+n,0}
\ee
We also define
\be
\a^{\dagger}_n = \a_{-n}
\ee
The vacuum $|0^k\rangle$ of the $k$-wound copy is defined by the condition
\be
\a_{n}|0^k\rangle = 0, ~~~~~~~n\geq 0 
\ee
The Virasoro generators can be expanded in terms of a sum over bilinears of modes
\be\label{4-7}
L_{m} = {1\over2k}\sum_n\a_{n}\a_{m-n}
\ee
again with implicit normal-ordering.
Using the commutation relation (\ref{comm twist}), we have
\be\label{L com}
[L_m,\a_n]=-n \a_{m+n}
\ee

The $k$-wound copy can be produced by applying the twist operator $\sigma_k$ to the untwisted sector. The twist operator $\sigma_k$ has dimension \cite{Lunin:2000yv}
\be\label{h}
h(\sigma_k)=\frac{c}{24}(k-\frac{1}{k})
\ee
For a single free boson, $c=1$.

\section{The effect of a twist operator}\label{sec ansatz}

In this section, we will briefly review the effect of a twist operator. In this paper, we restrict ourselves to the simplest case where there are only two copies of the seed CFT, such that $N=2$. 
Suppose at $z=0$ an initial state in the untwisted sector is given by
\be\label{i}
\a^{(i_1)}_{-n_1}\a^{(i_2)}_{-n_2}\dots \a^{(i_m)}_{-n_m}|0\rangle^{(1)}|0\rangle^{(2)}
\ee
where $n_k>0$ and $i_k=1,2$ is the copy label. Let us apply the twist operator $\sigma_2$ at $z_0$. The question is to find out the state $\phi$ at $|z|>|z_0|$ which is after the twist operator. The state $\phi$ is defined as
\be
|\phi\rangle=\sigma_2(z_0)\a^{(i_1)}_{-n_1}\a^{(i_2)}_{-n_2}\dots \a^{(i_k)}_{-n_k}|0\rangle^{(1)}|0\rangle^{(2)}
\ee
which lives on a doubly wound copy since the twist operator has joined the two singly wound copies in the initial state.
This question has been addressed completely by the covering map method, which will be reproduced for a single boson in appendix \ref{App cover}. The effect of a twist operator can be summarized by the following three basic rules:

\b

(i) Contraction: Two modes $\a^{(i)}_{-m}$ and $\a^{(j)}_{-n}$ in the initial state (\ref{i}) can `Wick contract', giving a number
\be
C^{ij}[m,n] \equiv C[\a^{(i)}_{-m}\a^{(j)}_{-n}] 
\ee
For the process of Wick contraction, we consider all possible pairs of modes. For each such pair, we get a term where the pair contracts to the above number, and a term where the pair does not contract but will pass through the twist as shown in step (ii) below. 

\b

(ii) Propagation: Any modes left after the contraction will pass through the twist and become modes after the twist operator, which are modes on the doubly wound copy.
\be\label{p rule}
\alpha^{(i)}_{-n}~~\longrightarrow~~  \sum_{p>0}f_i[-n,-p] \alpha_{-p},~~~~~i=1,2
\ee
where $\alpha_{-p}$ is a mode on the doubly wound copy and $p=m/2$ where $m$ is a positive integer. In section \ref{sec global}, we will show that the nontrivial $f_i$'s are
\bea\label{f mode}
f_i[-n,-n]&=&1/2 \nn
f_i[-n,-p]&\neq&  0,~~~~~\text{when}~~~ p\neq n ~~~\text{and}~~~ p ~\text{is a positive half integer}
\eea

\b

(iii) Pair creation: After the previous two steps, the modes in the initial state have either been contracted or passed through the twist. We are left with the twist operator acting on the untwisted vacuum
\bea\label{chi}
|\chi\rangle \equiv \s_2(z_0) |0\rangle^{(1)}|0\rangle^{(2)}=\text{exp}\big(\sum_{m,n>0}\gamma_{mn}\a_{-m}\a_{-n}\big)|0^2\rangle
\eea
where the dimension of the twist operator $\s_2$ is (\ref{h})
\be
h=h(\sigma_2)=1/16
\ee
which takes into account the difference of dimensions between the vacuum of doubly wound copy $|0^2\rangle$ and the vacuum of two singly wound copies $|0\rangle^{(1)}|0\rangle^{(2)}$.
In section \ref{sec global}, we will show that 
\be\label{gamma mode}
\gamma_{mn}\neq 0 ~~~~~ \text{only if $m,n$ are positive half integers}
\ee

\b 

\begin{figure}
\centering
        \includegraphics[width=10cm]{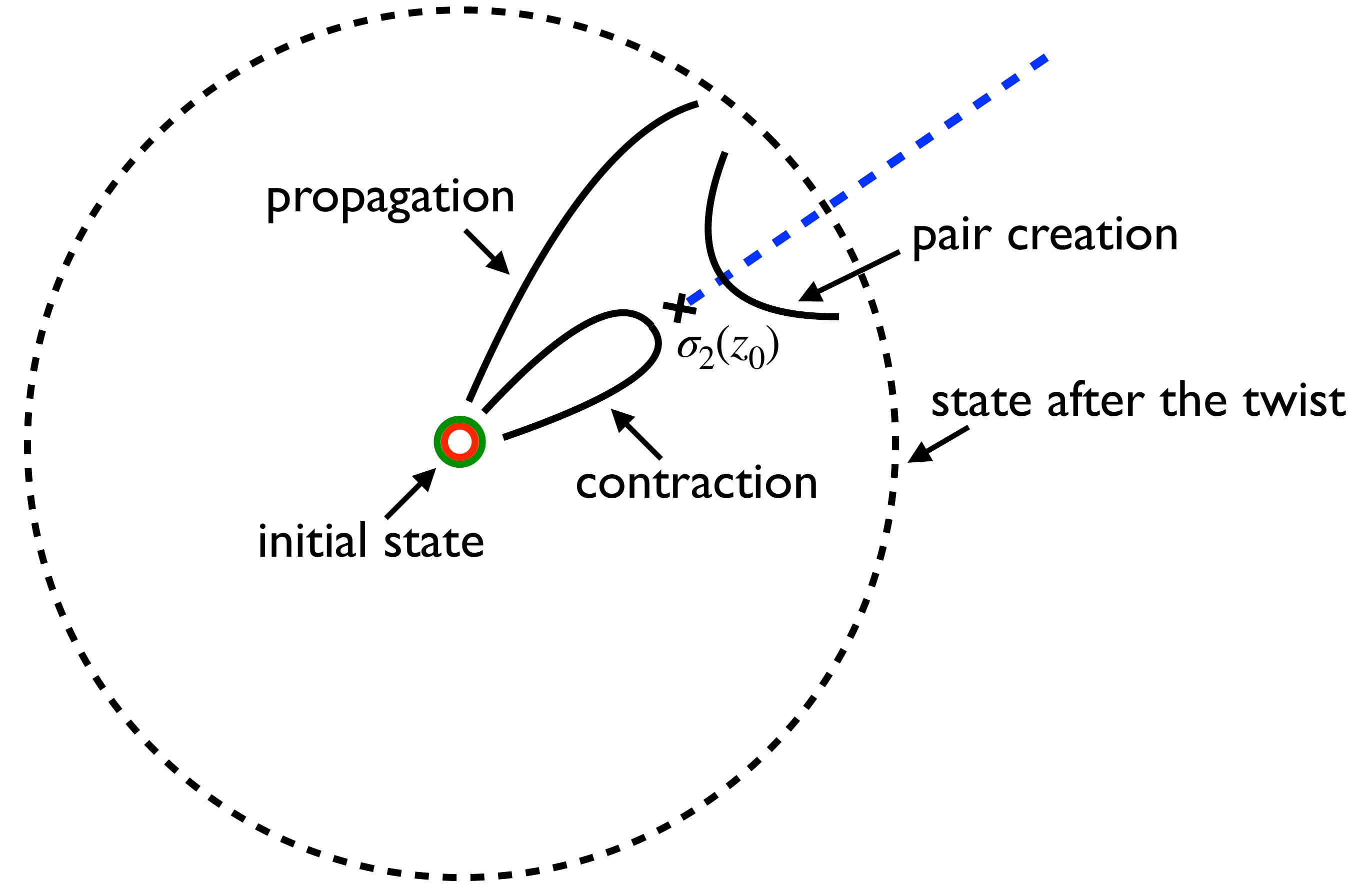}
\caption{The three basic rules to compute the effect of a twist operator.}
\label{fig_1}
\end{figure}

The above rules are shown in fig.\ref{fig_1}. 
To better understand these rules, let us consider an example with one initial mode
\bea\label{one}
\sigma_2(z_0)\a^{(i)}_{-n}|0\rangle^{(1)}|0\rangle^{(2)}
= \sum_{p>0}f_i[-n,-p]\a_{-p}
\, \text{exp}\big(\sum_{m,n>0}\gamma_{mn}\a_{-m}\a_{-n}\big)|0^2\rangle
\eea
We first use the propagation rule (\ref{p rule}) to pass the initial mode through the twist. Then the twist operator acts on the untwisted vacuum to produce pairs using the rule (\ref{chi}).
Let us now consider an example of two initial modes
\bea\label{two}
\sigma_2(z_0)\a^{(i)}_{-n_1}\a^{(j)}_{-n_2}|0\rangle^{(1)}|0\rangle^{(2)}
\!\!&=&\!\! \Big(\sum_{p_1>0}f_i[-n_1,-p_1]\a_{-p_1}\sum_{p_2>0}f_j[-n_2,-p_2]\a_{-p_2}+C^{ij}[n_1,n_2]\Big)\nn &&~\times\text{exp}\big(\sum_{m,n>0}\gamma_{mn}\a_{-m}\a_{-n}\big)|0^2\rangle
\eea
The first term in the parentheses comes from the propagation of the two initial modes while the second term is from the contraction. The exponent in the last line comes from the pair creation.

If the effect of an operator satisfies the above three rules but with independent and undetermined coefficients
$f_i$, $C^{ij}$, and $\gamma$, we call it the weak Bogoliubov form. As will be explained in the next section, in the `normal' Bogoliubov transformation these coefficients are not independent of each other.

\section{Bogoliubov transformation}\label{sec B}

The Bogoliubov transformation is a linear transformation that mixes creation and annihilation operators
\bea
\hat a&=&\a\,\hat b + \beta\,\hat b^\dagger \nn
\hat a^\dagger&=&\a^*\,\hat b^\dagger + \beta^*\,\hat b
\eea
where $|\a|^2-|\beta|^2 = 1$ to make both sets of operators canonical
\be
[\hat a,\hat a^\dagger]= 1,~~~~~~[\hat b,\hat b^\dagger]= 1
\ee
The `$a$' vacuum and `$b$' vacuum are defined by the conditions
\be
\hat a|0\rangle_a =0,~~~~~\hat b|0\rangle_b =0
\ee
Since the Bogoliubov transformation mixes the creation and annihilation operators, the `$a$' vacuum is no longer the `$b$' vacuum
\be
\hat a |0\rangle_a = (\a\,\hat b + \beta\,\hat b^\dagger) |0\rangle_a = 0
~~\longrightarrow~~ |0\rangle_a =  e^{\frac{1}{2}\gamma\, \hat b^\dagger \hat b^\dagger} |0\rangle_b
\ee
where
\be\label{b r}
\gamma = -\frac{\beta}{\alpha}
\ee
Consider an example with two initial modes on the `$a$' vacuum
\bea
\hat a^\dagger \hat a^\dagger |0\rangle_a 
&=& (\a^*\,\hat b^\dagger + \beta^*\,\hat b)(\a^*\,\hat b^\dagger + \beta^*\,\hat b)  e^{\frac{1}{2}\gamma \,\hat b^\dagger \hat b^\dagger} |0\rangle_b\nn
&=& \left[(\a^* + \beta^*\,\gamma)\hat b^\dagger (\a^* + \beta^*\,\gamma)\hat b^\dagger + \beta^*\alpha^* \right]  e^{\frac{1}{2}\gamma\, \hat b^\dagger \hat b^\dagger} |0\rangle_b
\eea
It is similar to the effect of a twist operator with two initial modes (\ref{two}).
The first term in the bracket comes from the propagation of the two initial modes while the second term is from their contraction. The exponent in the last line comes from the pair creation. Similarly, we can compute the case with multiple initial modes
\be
\hat a^\dagger\dots \hat a^\dagger|0\rangle_a
\ee
using the following three rules that are similar to the rules in section \ref{sec ansatz}. 

\b

(i) Contraction:
\be
C[\hat a^\dagger,\hat a^\dagger] = \beta^*\alpha^*
\ee

\b

(ii) Propagation: 

\be
\hat a^\dagger~ \rightarrow~ (\a^* + \beta^*\,\gamma)\hat b^\dagger
\ee

\b

(iii) Pair creation:

\be
\hat a |0\rangle_a = e^{\frac{1}{2}\gamma\, \hat b^\dagger \hat b^\dagger} |0\rangle_b
\ee

\b

The coefficients in these expressions are not independent of each other. They are related through (\ref{b r}).
We will call the above three rules with the constraint (\ref{b r}) the `normal' Bogoliubov ansatz and the same rules without the constraint the `weak' Bogoliubov ansatz.

In \cite{Carson:2014xwa}, it has been shown that the effect of a twist operator explained in section \ref{sec ansatz} satisfies the `normal' Bogoliubov ansatz with the generalization: the creation and annihilation operators are generalized to infinite-dimensional vectors, i.e. $\hat a$ becomes $\hat a_i$; the coefficients $\a$, $\beta$, $\gamma$ are generalized to infinite-dimensional matrices, i.e. $\a$ becomes $\a_{ij}$. In \cite{Carson:2014xwa}, by matching the modes before and after the twist, they obtained the matrices $\a$ and $\beta$. However, there is no systematic way to invert the infinite-dimensional matrices $\a$ to obtain $\gamma$ through the constraint (\ref{b r}).
In this paper, we will take the weak Bogoliubov ansatz which is without the constraint. 
We will show that it is enough to determine all the coefficients without inverting any infinite-dimensional matrices. The constraint (\ref{b r}) is satisfied automatically.

\section{Bootstrapping the effect of a twist operator}\label{bootstrap}

In this section, we will take the weak Bogoliubov ansatz and apply the Virasoro generators to obtain recursion relations for all the coefficients in the ansatz. The solutions to these recursion relations match the results from the covering map method. 

\subsection{Global modes}\label{sec global}

In this section we will show (\ref{f mode}) and (\ref{gamma mode}). Let us start with
\bea\label{global 1}
\alpha_{n}\sigma_2(z_0) - \sigma_2(z_0) (\alpha_{n}^{(1)}+\alpha_{n}^{(2)})
&=& \frac{1}{2\pi}\oint_{C_{z_0}^{(4\pi)}} dz~z^{n}\partial X(z)\sigma_2(z_0)\nn
&=& z_0^{n} \frac{1}{2\pi}\oint_{C_{z_0}^{(4\pi)}} dz~\big(1+(z-z_0)/z_0\big)^{n}\partial X(z)\sigma_2(z_0)\nn
&=& z_0^{n} \big(\alpha^{(z_0)}_0+n\alpha^{(z_0)}_{1}/z_0+\dots\big)\sigma_2(z_0)
\eea
where $n$ is an integer.
To obtain the first equality, we join the two contours, one before and one after the twist operator, into a single contour centered around the twist operator. 
We note that modes with superscript $(z_0)$, i.e. $\alpha^{(z_0)}_0$, are the modes centered around $z_0$, while modes without it, i.e. $\alpha_{n}$ and $\alpha_{n}^{(i)}$, are the modes centered around $z=0$. Since the twist operator $\sigma_2$ is defined as the lowest dimension operator that produces the twisted sector, we have
\be
\alpha^{(z_0)}_{n}\sigma_2(z_0)=0,~~~n\geq 0
\ee
Thus (\ref{global 1}) becomes
\be
\alpha_{n}\sigma_2(z_0) - \sigma_2(z_0) (\alpha_{n}^{(1)}+\alpha_{n}^{(2)}) = 0
\ee

Consider the following state where $m$ and $n$ are positive integers
\bea
\langle 0^2|\alpha_{m}\sigma_2(z_0) \alpha_{-n}^{(i)}|0\rangle^{(1)}|0\rangle^{(2)}
&=&\langle 0^2|\sigma_2(z_0)(\alpha^{(1)}_{m}+\alpha^{(2)}_{m}) \alpha_{-n}^{(i)}|0\rangle^{(1)}|0\rangle^{(2)}\nn
&=&\delta_{mn}n
\eea
which is nonzero only if $m$ and $n$ are equal.
From $\langle 0^2|\alpha_{n} \alpha_{-n}|0^2\rangle = 2n $, we find
\be
\sigma_2(z_0) \alpha_{-n}^{(i)}|0\rangle^{(1)}|0\rangle^{(2)} = \left(\frac{1}{2} \alpha_{-n}+ \text{half integer modes}\right)|0^2\rangle
\ee
which is (\ref{f mode}). To show (\ref{gamma mode}), consider the following state where $m$ is a positive integer
\be
\alpha_{m} |\chi\rangle=
\alpha_{m} \sigma_2(z_0) |0\rangle^{(1)}|0\rangle^{(2)}
=\sigma_2(z_0)(\alpha^{(1)}_{m}+\alpha^{(2)}_{m}) |0\rangle^{(1)}|0\rangle^{(2)} =0
\ee
Thus the state from the pair creation $\chi$ can not have any excitations of integer modes, which is stated in (\ref{gamma mode}).

\subsection{Pair creation}

In this subsection, we will derive the pair creation coefficients $\gamma_{m n}$.

\subsubsection{Relations from $L_{-1}$}

Starting with 
\be
L_{-1}|0\rangle^{(1)}|0\rangle^{(2)} = 0
\ee
where we have used (\ref{v untwist}) and (\ref{L untwist}), we have
\bea\label{4-5}
0&=&\s_2(z_0)L_{-1}|0\rangle^{(1)}|0\rangle^{(2)}\cr
&=&\big(L_{-1}\s_2(z_0)-[L_{-1},\s_2(z_0)]\big)|0\rangle^{(1)}|0\rangle^{(2)}
\eea
Let's compute the commutator. We have
\be\label{L-1 s}
[L_{-1},\s_2(z_0)]=\oint_{C^{(4\pi)}_{z_0}} {dz \over2\pi i}\,T(z)\s_2(z_0)=L^{(z_0)}_{-1} \s_2(z_0)
= \partial \s_2(z_0)
\ee
The mode $L^{(z_0)}_{-1}$ with superscript $(z_0)$ is the mode centered around $z_0$, while the mode without it, i.e. $L_{-1}$, is the mode centered around $z=0$.
Therefore (\ref{4-5}) becomes
\bea\label{pair L}
0&=&(L_{-1} - \partial)\s_2(z_0)|0\rangle^{(1)}|0\rangle^{(2)}\cr
&=&(L_{-1} - \partial)\text{exp}\big(\sum_{m,n\geq0}\gamma_{m+1/2,n+1/2}\,\a_{-(m+1/2)}\a_{-(n+1/2)}\big)|0^2\rangle
\eea
Here we have relabeled $m\rightarrow m+1/2$ and $n\rightarrow n+1/2$ in (\ref{chi}) where $m$ and $n$ are now integers.
Our aim is to find relations which will help us solve for $\gamma_{m+1/2,n+1/2}$. 
There are four cases we will look at it.

\subsubsection*{\underline{$m,n>0$}}

The first case is $m,n>0$. We will look for ways to obtain terms propotional to 
\be
\a_{-(m+1/2)}\a_{-(n+1/2)}|0^2\rangle
\ee
There are three ways to get this. From (\ref{pair L}) we obtain
\bea
0\!\!&=&\!\!\gamma_{m-1/2,n+1/2}\big[L_{-1},\a_{-(m-1/2)}\big]\a_{-(n+1/2)} + \gamma_{m+1/2,n-1/2} \a_{-(m+1/2)}\big[L_{-1},\a_{-(n+1/2)}\big]\cr
&&-\partial\, \gamma_{m+1/2,n+1/2}\a_{-(m+1/2)}\a_{-(n+1/2)}
\eea 
Using (\ref{4-9}) we find the relation
\be\label{gamma e}
\gamma_{m-1/2,n+1/2}(m-1/2) + \gamma_{m+1/2,n-1/2}(n-1/2) = \partial\, \gamma_{m+1/2,n+1/2}
\ee

\subsubsection*{\underline{$m=0,n>0$}}

Next we consider the case where $m=0,n>0$. We need terms proportional to 
\be
\a_{-1/2}\a_{-(n+1/2)}|0^2\rangle
\ee
Looking at (\ref{pair L}) we get
\be
0=\gamma_{1/2,n-1/2}\a_{-1/2}[L_{-1},\a_{-(n-1/2)}] -\partial\, \gamma_{1/2,n+1/2}\a_{-1/2}\a_{-(n+1/2)}
\ee
Again using (\ref{4-9}) we obtain the relation
\be\label{4-18}
\gamma_{1/2,n-1/2}(n-1/2)=\partial\,\gamma_{1/2,n+1/2}
\ee

\subsubsection*{\underline{$m>0,n=0$}}

Similar to the previous case, for $m>0,n=0$ we obtain the relation
\be \label{4-19}
\gamma_{m-1/2,1/2}(m-1/2)=\partial\,\gamma_{m+1/2,1/2}
\ee

\subsubsection*{\underline{$m=n=0$}}

For $m=n=0$ we need terms proportional to 
\be
\a_{-1/2}\a_{-1/2}|0^2\rangle
\ee
For this we use (\ref{4-7}) with $k=2$ where
\be
L_{-1} |0^2\rangle={1\over 4}\a_{-1/2}\a_{-1/2}|0^2\rangle
\ee
Using this in (\ref{pair L}) we find the relation
\be \label{gamma 1/2}
{1\over 4}=\partial\,\gamma_{1/2,1/2}
\ee

\subsubsection{The solution}

To find the initial condition for these differential equations, we consider
\bea
|0^2\rangle &=& \sigma_2 (z_0=0) |0\rangle^{(1)}|0\rangle^{(2)} \nn
&=& \text{exp}\big(\sum_{m,n\geq0}\gamma_{m+1/2,n+1/2}(z_0=0)\,\a_{-(m+1/2)}\a_{-(n+1/2)}\big)|0^2\rangle
\eea
The first line comes from the definition of the twist operator $\sigma_2$ which is the lowest dimension operator that changes the untwisted sector to the twisted sector. Thus we have
\be
\gamma_{m+1/2,n+1/2}(z_0=0) = 0
\ee
Using these initial conditions, we can solve the differential equations. The solution to (\ref{gamma 1/2}) is
\be
\gamma_{1/2,1/2} = {1\over 4}z_0
\ee
We can find all other $\gamma_{m+1/2,n+1/2}$'s by using the relations (\ref{gamma e}), (\ref{4-18}), and (\ref{4-19}). You can think of the $\gamma_{m+1/2,n+1/2}$'s as forming an inverted triangle with integer lattice spacing with $\gamma_{1/2,1/2}$ as the bottom lattice point. The relation (\ref{4-18}) moves you along the right edge, (\ref{4-19}) moves you along the left edge and (\ref{gamma e}) moves you within the interior.
\be
\begin{tikzcd}[row sep=16pt, column sep=-5pt]
~~~~~~& &~~~~~~~~ & &~~~~~~~~ & & ~~~~~~ \\
 &\arrow[ul,dotted]\gamma_{1/2,5/2}\arrow[ur,dotted]& &\arrow[ul,dotted]\gamma_{3/2,3/2} \arrow[ur,dotted] & &\arrow[ul,dotted]\gamma_{5/2,1/2}\arrow[ur,dotted] & \\
 & &\arrow[ul]\gamma_{1/2,3/2}\arrow[ur] &  &\arrow[ul]\gamma_{3/2,1/2}\arrow[ur] & &\\
 & & & \arrow[ul]\gamma_{1/2,1/2} \arrow[ur]& & &
\end{tikzcd}
\ee
The solution is
\be\label{gamma}
\gamma_{m+1/2,n+1/2}=\frac{z_0^{m+n+1}\Gamma[\frac{3}{2}+m]\Gamma[\frac{3}{2}+n]}{(2m+1)(2n+1)(1+m+n)\pi \Gamma[m+1]\Gamma[n+1]}
\ee
where $m$ and $n$ are non-negative integers. 

\subsection{Propagation}\label{sec f}

Here we derive the propagation coefficients $f_i[-n,-p]$ which correspond to a mode passing through the twist.

\subsubsection{Relations from $L_0$}

Here we use the generator $L_0$ to derive a relation for $f_1[-1,-p]$. We begin with 
\bea
\s_2(z_0) \a^{(1)}_{-1}|0\rangle^{(1)}|0\rangle^{(2)}&=&\s_2(z_0) L_0\a^{(1)}_{-1}|0\rangle^{(1)}|0\rangle^{(2)}\nn
&=& (L_0\s_2(z_0) - [L_0,\s_2(z_0)]) \a^{(1)}_{-1}|0\rangle^{(1)}|0\rangle^{(2)}
\eea
where the initial mode is on the first copy.
Let us compute the commutator. We have
\bea\label{L0}
[L_0,\s_2(z_0)] &=& \oint_{C^{(4\pi)}_{z_0}} {dz\over2\pi i}\,z\,T(z)\s_2(z_0)\cr
&=& \oint_{C^{(4\pi)}_{z_0}} {dz\over2\pi i}\,(z_0+z-z_0)\,T(z)\s_2(z_0)\cr
&=& (z_0L^{(z_0)}_{-1} + L^{(z_0)}_{0})\s_2(z_0)\cr
&=&(z_0\partial+h) \s_2(z_0)
\eea 
Therefore
\be
\s_2(z_0) \a^{(1)}_{-1}|0\rangle^{(1)}|0\rangle^{(2)}= (L_0 - (z_0\partial+h))\s_2(z_0) \a^{(1)}_{-1}|0\rangle^{(1)}|0\rangle^{(2)}
\ee
which gives
\be
\sum_{p> 0}f_1[-1,-p]\a_{-p}|\chi\rangle
=(L_0 - (z_0\partial+h))\sum_{p> 0}f_1[-1,-p]\a_{-p}|\chi\rangle
\ee
where $\chi$ is state (\ref{chi}) from pair creation. Taking the term proportional to $\a_{-p}|0^2\rangle$ with $p$ being a positive half integer, we have
\be
f_1[-1,-p]\a_{-p}|0^2\rangle
=(L_0 - (z_0\partial+h))f_1[-1,-p]\a_{-p}|0^2\rangle
\ee
which gives
\be
f_1[-1,-p]
=(p - z_0\partial)f_1[-1,-p]
\ee
The solution is
\be\label{f pro}
f_1[-1,-p]\propto z_0^{p-1}
\ee
where $p$ is a positive half integer.

\subsubsection{Relations from $L_1$}

Here we use $L_1$ to determine the proportional coefficients in (\ref{f pro}). 
Since
\be
\a^{(1)}_{0}|0\rangle^{(1)} = 0
\ee
we begin with the following relation
\be
0=\s_2(z_0) L_1\a^{(1)}_{-1}|0\rangle^{(1)}|0\rangle^{(2)}
\ee
Bringing $L_1$ through the twist we obtain
\be \label{4-25}
0= (L_1\s_2(z_0) - [L_1,\s_2(z_0)]) \a^{(1)}_{-1}|0\rangle^{(1)}|0\rangle^{(2)}
\ee
Let us compute the commutator. We have
\bea
[L_1,\s_2(z_0)] &=& \oint_{C^{(4\pi)}_{z_0}} {dz\over2\pi i}\,z^{2}\,T(z)\s_2(z_0)\cr
&=& \oint_{C^{(4\pi)}_{z_0}} {dz\over2\pi i}\,(z_0+z-z_0)^{2}T(z)\s_2(z_0)\cr
&=&(z_0^2 L^{(z_0)}_{-1}+2z_0 L^{(z_0)}_0 + L^{(z_0)}_1)\s_2(z_0)\cr
&=&z_0(z_0\partial + 2 h)\s_2(z_0)
\eea 
Inserting this into (\ref{4-25}) yields
\bea \label{4-28}
0&=& (L_1 -z_0(z_0\partial + 2 h))\s_2(z_0) \a^{(1)}_{-1}|0\rangle^{(1)}|0\rangle^{(2)}\cr
&=& (L_1 -z_0(z_0\partial + 2 h))\sum_{p> 0}f_1[-1,-p]\a_{-p}\text{exp}\big(\sum_{m,n\geq0}\gamma_{m+1/2,n+1/2}\a_{-(m+1/2)}\a_{-(n+1/2)}\big)|0^2\rangle\nn
\eea 
We again match terms to obtain relations which one solves to find $f_1[-1,-p]$. We want to keep terms which are $\a_{-(m+1/2)}|0^2\rangle$ with $m>0$. The terms which contribute are
\bea 
0&=&2f_1[-1,-1/2]\gamma_{m+1/2,1/2}\a_{-(m+1/2)}[[L_1,\a_{-1/2}],\a_{-1/2}]\cr
&& + f_1[-1,-(m+1/2)]\gamma_{1/2,1/2}\a_{-(m+1/2)}[[L_1,\a_{-1/2}],\a_{-1/2}]\cr
&&+f_1[-1,-(m+3/2)][L_1,\a_{-(m+3/2)}]-z_0(z_0\partial + 2 h)f_1[-1,-(m+1/2)] \a_{-(m+1/2)}\nn
\eea 
Using the commutation relation (\ref{L com}), this gives a recursion relation for $m>0$
\bea\label{f re1}
&&(m+3/2)f_1[-1,-(m+3/2)]\nn
&=&[z_0(z_0\partial + 2 h) -{1\over2} \gamma_{1/2,1/2}]f_1[-1,-(m+1/2)]-f_1[-1,-1/2]\gamma_{m+1/2,1/2} 
\eea
Now we need to find a relation for $f_1[-1,-3/2]$. In order to do so we will need terms which are proportional to $\a_{-1/2}|0^2\rangle$. From (\ref{4-28}) we get
\bea 
0&=&3f_1[-1,-1/2]\gamma_{1/2,1/2}[[L_1,\a_{-1/2}],\a_{-1/2}]\a_{-1/2}\nn
&&+f_1[-1,-3/2][L_1,\a_{-3/2}]-z_0(z_0\partial + 2 h)f_1[-1,-1/2] \a_{-1/2}
\eea 
where we have collected terms coming from appropriate commutators in order to leave only one $\a_{-1/2}$. Again using commutation relations we obtain the relation
\bea\label{f re2}
{3\over2}f_1[-1,-3/2]=z_0(z_0\partial + 2 h)f_1[-1,-1/2]-{3\over2}f_1[-1,-1/2]\gamma_{1/2,1/2}
\eea
Therefore all $f_1[-1,-p]$ can be determined from $f_1[-1,-1/2]$.

\subsubsection{The solution}\label{sec f s}

The solution satisfying the recursion relations (\ref{f re1}) and (\ref{f re2}) is
\be\label{f1pc1}
f_1[-1,-p]=C\frac{\Gamma[p-1]}{\Gamma[p+1/2]}z_0^{p-1}
\ee
where $p$ is a positive half integer.
The constant $C$ will be derived in (\ref{C})
\be\label{C 0}
C=\pm \frac{i}{4\sqrt{\pi}}
\ee
To obtain $f_2[-1,-p]$ for an excitation on copy 2, we change the location of the twist by $z_0\to z_0 e^{2\pi i}$. It interchanges copy 1 and copy 2. Since $p$ is a half integer, we have $z_0^{p-1} \to -z_0^{p-1}$. Therefore, we have
\be\label{f2pc1}
f_2[-1,-p]=-C\frac{\Gamma[p-1]}{\Gamma[p+1/2]}z_0^{p-1}
\ee
Thus the two possible signs of (\ref{C 0}) correspond to two possible conventions of labeling copy 1 and copy 2. By applying $L_{-1}$ repeatedly, we can compute $f_i[-n,-p]$ for $n>1$, which will be shown in appendix \ref{App f}. The final expressions are given by
\bea\label{final f int}
f_1[-n,-p]&=&{1\over2} \delta_{n,p}\cr
f_1[-n,-p]
&=&{iz_0^{p-n}\Gamma({1\over2}+n)\Gamma(p)\over\pi(2p - 2n)\Gamma(n)\Gamma(p+\frac{1}{2})},\quad n\neq p
\cr
f_2[-n,-p]&=&{1\over2} \delta_{n,p}\cr
f_2[-n,-p]&=&-f_1[-n,-p],\quad n\neq p
\eea

\subsection{Contraction}

Here we derive the expression for the contraction of two modes in the initial state under the effect of the twist.

\subsubsection{Relations from $L_{1}$}

Let's start with the following state with $n>1$
\bea
\s(z_0)\a^{(i)}_{-1}\a^{(j)}_{-(n-1)}|0\rangle^{(1)}|0\rangle^{(2)}&=&\frac{1}{n}
\s(z_0)L_1\a^{(i)}_{-1}\a^{(j)}_{-n}|0\rangle^{(1)}|0\rangle^{(2)}\cr
&=&{1\over n}\big[L_1-z_0(z_0\partial + 2 h)\big]\s_2(z_0)\a^{(i)}_{-1}\a^{(j)}_{-n}|0\rangle^{(1)}|0\rangle^{(2)}
\eea 
where we remind the reader that $i,j=1,2$ are copy labels.
So we have the relation
\bea\label{C r1}
[L_1-z_0(z_0\partial + 2 h)]\s_2(z_0)\a^{(i)}_{-1}\a^{(j)}_{-n}|0\rangle^{(1)}|0\rangle^{(2)} = n\s(z_0)\a^{(i)}_{-1}\a^{(j)}_{-(n-1)}|0\rangle^{(1)}|0\rangle^{(2)}
\eea 
which becomes
\bea
&&[L_1-z_0(z_0\partial + 2 h)]\Big(\sum_{p> 0}f_i[-1,-p]\a_{-p}\sum_{p'> 0}f_j[-n,-p']\a_{-p'} + C^{ij}[1,n]\Big)|\chi\rangle\cr
&=& n\Big(\sum_{p> 0}f_i[-1,-p]\a_{-p}\sum_{p'> 0}f_j[-(n-1),-p']\a_{-p'} +C^{ij}[1,n-1]\Big)|\chi\rangle
\eea 
We collect terms which carry no bosonic modes in order to easily isolate $C^{ij}$. Terms of this kind are given by
\bea
&&\big(f_i[-1,-1/2]f_j[-n,-1/2]+C^{ij}[1,n]\gamma_{1/2,1/2} \big)[[L_1,\a_{-1/2}],\a_{-1/2}]\cr
&&-z_0(z_0\partial + 2 h)C^{ij}[1,n]=nC^{ij}[1,n-1]
\eea
Using the commutation relations yield for $n>1$
\be 
{1\over2}f_i[-1,-1/2]f_j[-n,-1/2]+\frac{z_0}{8}C^{ij}[1,n] -z_0(z_0\partial + 2 h)C^{ij}[1,n]=nC^{ij}[1,n-1]
\ee
Setting $h=1/16$ (the dimension of the twist operator), the relation becomes
\be \label{C r2}
{1\over2}f_i[-1,-1/2]f_j[-n,-1/2] -z_0^2 \partial C^{ij}[1,n]=nC^{ij}[1,n-1]
\ee
where $n>1$.
To find $C^{ij}[1,1]$, notice that in (\ref{C r1}) if $n=1$, the RHS vanishes. Thus the RHS of (\ref{C r2}) vanishes if $n=1$, which gives
\be
{1\over2}f_i[-1,-1/2]f_j[-1,-1/2]=z_0^2\partial C^{ij}[1,1]
\ee
The solution is
\be\label{C11}
C^{ij}[1,1] =
-(-1)^{i+j}\pi C^2 z_0^{-2}
\ee
where we have used (\ref{f1pc1}) and (\ref{f2pc1}).

\subsubsection{Relations from $L_{-2}$}

Here we will derive the constant $C$ which appears in (\ref{f1pc1}), (\ref{f2pc1}), and (\ref{C11}).
Start with 
\be\label{C 1}
\sigma (z_0)L_{-2} |0\rangle^{(1)}|0\rangle^{(2)} 
=\sigma (z_0)\frac{1}{2}\big(\alpha^{(1)}_{-1}\alpha^{(1)}_{-1}+\alpha^{(2)}_{-1}\alpha^{(2)}_{-1}\big)|0\rangle^{(1)}|0\rangle^{(2)}
\ee
Using the contraction (\ref{C11}), the term without any modes on the RHS is
\be\label{R C}
\frac{1}{2}\big(C^{11}[1,1]+C^{22}[1,1]\big)|0^2\rangle = - \pi C^2 z_0^{-2}|0^2\rangle
\ee
Let us compute the LHS of (\ref{C 1}) in another way. We have
\be
\s_2(z_0)L_{-2}=L_{-2}\s_2(z_0) - [L_{-2},\s_2(z_0)] 
\ee
where
\bea
[L_{-2},\s_2(z_0)] &=& \oint_{C^{(4\pi)}_{z_0}} {dz\over2\pi i}z^{-1}T(z)\s_2(z_0)\cr
&=& z_0^{-1}\oint_{C^{(4\pi)}_{z_0}} {dz\over2\pi i} \big(1+\frac{z-z_0}{z_0}\big)^{-1}T(z)\s_2(z_0)\cr
&=& z_0^{-1}\oint_{C^{(4\pi)}_{z_0}} {dz\over2\pi i} \big(1-\frac{z-z_0}{z_0}+\dots\big)T(z)\s_2(z_0)\cr
&=& z_0^{-1}\big(L^{(z_0)}_{-1}-z_0^{-1}L^{(z_0)}_{0}\big)\s_2(z_0)\cr
&=& z_0^{-1}\big(\partial-z_0^{-1}h\big)\s_2(z_0)
\eea
Thus we have
\be
\sigma (z_0)L_{-2} |0\rangle^{(1)}|0\rangle^{(2)}\nn
=\big[L_{-2}-z_0^{-1}(\partial-z_0^{-1}h)\big]|\chi\rangle
\ee
The term without any modes is
\be
 z_0^{-2} h |0^2\rangle = \frac{z_0^{-2}}{16} |0^2\rangle
\ee
where we have used $h=1/16$.
Comparing to (\ref{R C}), we obtain
\be
- \pi C^2 = \frac{1}{16}
\ee
which gives
\be\label{C}
C=\pm \frac{i}{4\sqrt{\pi}}
\ee
As explained in section \ref{sec f s}, the two signs correspond to the two different conventions of labeling the copies.

\subsubsection{The solution}

Using (\ref{C}), the contraction (\ref{C11}) becomes
\be\label{C11 f}
C^{ij}[1,1] =
-(-1)^{i+j}\pi C^2 z_0^{-2}=(-1)^{i+j}{1\over16}z_0^{-2}
\ee
Solving (\ref{C r2}) recursively we find the following solution
\be
C^{ij}[1,n] = (-1)^{i+j}{z_0^{-(1+n)}\Gamma({1\over2}+n)\over4(1+n)\sqrt\pi\Gamma(n)} \label{C111n}
\ee
The contractions $C^{ij}[n_1,n_2]$ with $n_1,n_2>0$ are computed in appendix \ref{Cij}. The final expressions are 
\be\label{final Cij}
C^{ij}[n_1,n_2]
= (-1)^{i+j}{z_0^{-(n_1+n_2)}\over2(n_1+n_2)\pi}{\Gamma({1\over2}+n_1)\Gamma({1\over2}+n_2)\over\Gamma(n_1)\Gamma(n_2)}
\ee
where $i,j=1,2$ are copy labels.

\section{Discussion}\label{discussion}

The traditional way to compute the effect of a twist operator is by using the covering map method. 
In this paper, we have developed a new method using the Bogoliubov ansatz and conformal symmetry. 
The Bogoliubov ansatz includes three quantities which characterize the final state uniquely. These three quantities correspond to three effects produced by the twist operator. One effect is pair creation in which the twist, when acting on the vacuum in the initial state, produces pairs of modes in the final state. This effect is encoded in the coefficients $\gamma_{mn}$ which are computed in (\ref{gamma}). Another effect, which arises by applying the twist operator to a single mode in the initial state, is propagation. This effect is encoded in the functions $f_i[-n,-p]$ which are computed in (\ref{final f int}). 
The third effect produced by the twist is the contraction of two modes in the initial state. This process is encoded in the functions $C^{ij}[n_1,n_2]$. These expressions were computed in (\ref{final Cij}). Each of these quantities agrees with the results from the covering map method in appendix \ref{App cover}. Using the Bogoliubov ansatz along with conformal symmetry we have derived a new method for computing effects of the twist operator in orbifold CFTs.

Our results also answer an important question about the nature of the effect of a twist operator. 
From the general idea of the covering map, we know that the twist operator generates a Bogoliubov transformation.  This raises the following question: if we know that the effect of a twist is a Bogoliubov transformation what else do we need to completely determine the coefficients in this transformation? 
From the covering map method, it seems that the answer is the covering map. The covering map is essential to determine the effect of a twist operator. However, our results show that with conformal symmetry we do not need other input to determine its effect. The nature of the effect of a twist operator is captured by the form of the Bogoliubov transformation and conformal symmetry.

In the study of the perturbative D1D5 CFT, twist operators and their effects carry a wide variety of applications since a class of marginal deformations of the theory contain twist operators \cite{Gava:2002xb,Gaberdiel:2015uca,Burrington:2018upk,Hampton:2018ygz,DeBeer:2019oxm,Guo:2019pzk,Keller:2019yrr,Hampton:2019hya,Hampton:2019csz,Guo:2019ady,Guo:2020gxm,Lima:2020boh,Lima:2020nnx,Lima:2020urq,Lima:2020kek,Lima:2021wrz,Benjamin:2021zkn,AlvesLima:2022elo,Apolo:2022fya}. 
The D1D5 CFT is realized by four free bosons and four free fermions. 
Our results, which are for one boson, can be simply generalized to four bosons where each boson has its own propagation, contraction and pair creation. These effects for each of the four bosons of the D1D5 CFT are the same as those of one free boson computed in this paper, thus yielding the same expressions for the three relevant quantities. 
The generalization to free fermions is straight forward but will contain some additional features. We will present these results in a future work. Furthermore, to compute higher order effects in the D1D5 CFT, which is relevant for holography, it is necessary to compute higher order twist correlation functions \cite{Carson:2016cjj,Carson:2016uwf,Carson:2015ohj,Guo:2021ybz,Guo:2021gqd}. In the covering map method, these computations become more challenging due to the growing complexity of the covering maps themselves. The method developed in this paper provides tools to possibly compute these higher order effects.  

\section*{Acknowledgements}
We would like to thank Soumangsu Chakraborty, Nicolas Kovensky, Samir Mathur and Hynek Paul for helpful discussions. The work of B.G. and S.H. is supported ERC Grant 787320 - QBH Structure.

\appendix

\section{Covering space method}\label{App cover}

In this appendix we review the covering space method which is traditionally used to compute the effect of a twist operator. We will compute the coefficients of the three rules in section \ref{sec ansatz}. In \cite{Avery:2010er,Avery:2010hs}, these coefficients are computed for the D1D5 CFT with $\mathcal{N}=4$ superconformal symmetry, which is realized by four free bosons and four free fermions. In this appendix, we will focus on the theory of one free boson. The results are similar to the boson in the supersymmetric case with some normalization differences.

\subsection{Mode Definitions}

Let us first define the bosonic modes on the $z$-plane as in section \ref{sec mode}. Imagine we place the twist operator at a point $z_0$. Before the twist which corresponds to the point, $|z|<|z_0|$, we have two singly wound copies. They are defined as follows
\bea
\a^{(i)}_{m} = {1\over2\pi}\oint_{C_0}dzz^{m}\partial X^{(i)} (z),\quad i=1,2
\eea 
Here $m$ is an integer.
The commutation relations before the twist is given by 
\bea
[\a^{(i)}_{m},\a^{(j)}_{n}]=m\,\d^{ij}\d_{m+n,0}
\eea
After the twist where $|z| > |z_0|$, we have a doubly wound copy and the modes are defined as follows
\bea
\a_{m} = {1\over2\pi}\oint_{C^{(4\pi)}_0}dzz^m\partial X (z)
\eea
Here $m$ can be integer or half integer since these are modes defined on a doubly wound copy (we will indeed show that only half integer modes are nontrivially affected by the twist operator).
The commutation relation after the twist is given by 
\bea\label{comm after twist}
[\a_{m},\a_{n}]=2m\d_{m+n,0}
\eea
where the factor of two comes from the fact that the theory is defined on a doubly wound copy. Next we discuss the covering map.
\subsection{Covering Map}
In order to compute these quantities we use the covering space method. In the $z$-plane, the effect of the twist introduces a branch cut. To resolve this branch cut, one can pass to the covering space $t$ defined as
\be\label{cover map}
z = z_0 + t^2
\ee
where 
\be
z_0=a^2
\ee
For $|z|>|z_0|$ we have a doubly wound copy. In the $t$-plane, $|z|\to\infty$ corresponds to $t\to\infty$.
In the $z$-plane we have two single copies in the initial state located at the origin, $z=0$. The location of these states gets mapped to two different points in the $t$-plane. In our convention, they are
\bea 
\text{Copy 1}:\quad t=ia,\quad
\text{Copy 2}:\quad t=-ia
\eea 
One can also use the other convention where the copy labels are interchanged.
The location of the twist operator is at $z_0=a^2$. It maps to the location $t=0$ in the $t$-plane.
Here the covering map is of order $2$ and is therefore the location where one crosses from one branch to the other. 
Now that we have analyzed the covering map let us compute the various ansatz quantities by mapping them to the $t$-plane. In the following sections we will compute the quantities, $\gamma,f_i,C^{ij}$, which determine the twisted state in the theory of one free boson.

\subsection{Pair creation}

In this subsection we compute the Bogoliubov coefficient $\gamma$ by using the covering map (\ref{cover map}). We start with the state (\ref{chi}). Let us first compute the amplitude 
\bea 
\langle0^2|\a_{m}\a_{n}\s_2(z_0)|0\rangle^{(1)}|0\rangle^{(2)}&=&\langle0^2|\a_{m}\a_{n}|\chi\rangle
\eea 
Using the expression (\ref{chi}) we obtain the following
\bea
\gamma_{mn} = {1\over8mn}\langle0^2|\a_{m}\a_{n}\s_2(z_0)|0\rangle^{(1)}|0\rangle^{(2)} 
\eea 

In order to remove the twist we map our result to the $t$-plane. This gives the relation
\bea\label{gamma t}
\gamma_{mn} &=& {1\over8mn}\,{}_t\langle0|\a'_{m}\a'_{n}|0\rangle_t
\eea
The primes denote modes which have been mapped to the $t$-plane and the subscript $t$ denotes $t$-plane states. The state $|0\rangle_t$ is the vacuum at the origin and the conjugate state ${}_t\langle 0 |$ is the vacuum at $t=\infty$. We see that the insertion of the twist is removed since it's action is encoded in the covering map. The problem is then simplified to computing a wick contraction between terms within the $t$-plane. To do so we must define modes which are natural to the $t$-plane. For $\gamma_{mn}$ we will only need modes defined after the twist but for the other coefficients we'll need modes before the twist. We therefore record both types below.

\subsubsection*{Modes before the twist}

Modes natural to the $t$-plane which correspond to initial states are given by
\bea \label{nat t init}
\text{Copy 1: } \tilde\a^{t\to ia}_{m} &=& {1\over2\pi}\oint_{C_{ia}}dt(t-ia)^m\partial X(t)\cr
\text{Copy 2: } \tilde\a^{t\to -ia}_{m} &=& {1\over2\pi}\oint_{C_{-ia}}dt(t+ia)^m\partial X(t)
\eea
Here $m$ is an integer. We have the commutation relations
\bea\label{init comm t}
[\tilde\a^{t\to ia}_{m},\tilde\a^{t\to ia}_{n}] &=& m\d_{m+n,0}\cr
[\tilde\a^{t\to -ia}_{m},\tilde\a^{t\to -ia}_{n}] &=& m\d_{m+n,0}
\eea

\subsubsection*{Modes after the twist}

Modes natural to the $t$-plane corresponding to the image of the location of final states in the $z$-plane are given by
\bea\label{boson t}
\tilde\a^{t\to\infty}_{m} = {1\over2\pi}\oint_{C_\infty}dt \,t^m\partial X(t)
\eea
Here $m$ is an integer. Also we have the commutation relation
\bea\label{inf comm t}
[\tilde\a^{t\to\infty}_{m},\tilde\a^{t\to\infty}_{n}] = m\d_{m+n,0}
\eea
Modes defined after the twist in the $z$-plane map to 
\be
\label{boson prime}
\a_m\to \a'_m = {1\over2\pi}\oint_{C_\infty} dt(z_0 +t^2)^{m}\partial X(t)
\ee
Let's expand this mode around $t=\infty$. The integrand is 
\be
(z_0 +t^2)^m = t^{2m}(1 + z_0t^{-2})^{m}
=\sum_{k\geq0}{}^{m}C_k\,z_0^{k}\,t^{2m-2k}
\ee
Inserting this expansion into (\ref{boson prime}) and using (\ref{boson t}) gives
\be\label{prime infty}
\a'_m = \sum_{k\geq0}{}^{m}C_k\,z_0^{k}\,\tilde\a^{t\to\infty}_{2m-2k}
\ee
Inserting this into (\ref{gamma t}) gives 
\be
\gamma_{mn} = {1\over8mn} \sum_{k,k'\geq0}{}^{m}C_k{}^{n}C_{k'}z_0^{k+k'}[\tilde\a^{t\to\infty}_{2m-2k},\tilde\a^{t\to\infty}_{2n-2k'}]
\ee
In order to have a nonzero contraction we require
\bea 
m>k \text{  and  } n<k'
\eea
Using commutation relation (\ref{inf comm t}) gives
\be
\gamma_{mn} = {1\over4mn}\sum_{k=0}^{\lceil m -1\rceil}(m-k)\,{}^{m}C_k\,{}^{n}C_{m+n-k}\,z_0^{m+n}
\ee
For $m,n\in\mathbb{Z}$, $\gamma_{mn}$ is zero. For the case where $m,n$ are half-integers we take 
\bea
m&=&m'+1/2\cr
n&=&n'+1/2
\eea 
and we obtain
\be
\gamma_{m'+1/2,n'+1/2}={z_0^{m'+n'+1}\Gamma({3\over2}+m')\Gamma({3\over2}+n')\over(2m'+1)(2n'+1)(1+m'+n')\pi\Gamma(1+m')\Gamma(1+n')}
\ee
where $m',n'\in \mathbb{Z}_+$.

\subsection{Propagation}

In this subsection we compute $f_i[-n,-p]$ which describes the propagation of a mode through the twist operator where $i=1,2$ is the copy label of the initial mode. Let us start with the state 
\be
 \s_2(z_0) \a^{(i)}_{-n}|0\rangle^{(1)}|0\rangle^{(2)}
=\sum_{p> 0}f_i[-n,-p]\a_{-p}|\chi\rangle
\ee
The propagation $f_i$ can be computed from
\be
f_i[-n,-p]=  {1\over2p}\langle 0^2|\a_{p}\s(z_0)\a^{(i)}_{-n}|0\rangle^{(1)}|0\rangle^{(2)}
\ee
where we have used the commutation relation (\ref{comm after twist}).

Again, mapping to the $t$-plane gives
\be\label{fi}
f_i[-n,-p] =  {1\over2p}{}_t\langle 0|\a'_{p}\a'^{(i)}_{-n}|0\rangle_t
\ee
In order to compute the Wick contraction we will first expand the mode $\a'^{(i)}_{-n}$ around it's $t$-plane image corresponding to a mode in the initial state. We will then expand these modes at $t=\infty$ enabling us to perform the Wick contraction with the mode in the final state. We first write the mode $\a'^{(1)}_{-n}$ which maps to the image $t=ia$
\bea\label{ini1 t}
\a'^{(1)}_{-n} = {1\over2\pi}\oint_{C_{ia}} dt (a^2 + t^2)^{-n}\partial X(t)
\eea
We expand the integrand around $t=ia$ as follows
\bea
(a^2 + t^2  )^{-n}
&=&(t-ia)^{-n}(2ia + t - ia )^{-n}\cr
&=&\sum_{k\geq0}{}^{-n}C_k\,(2ia)^{-(n+k)}(t-ia)^{k-n}
\eea
Inserting this expansion into (\ref{ini1 t}) and using the definition in (\ref{nat t init}) gives
\bea 
\a'^{(1)}_{-n} = \sum_{k\geq0}{}^{-n}C_k\,(2ia)^{-(n+k)}\tilde\a^{t\to ia}_{k-n}
\eea
We note that for copy 2 we simply take $a\to - a$ which gives the expansion 
\bea\label{boson prime initial 2} 
\a'^{(2)}_{-n} = \sum_{k\geq0}{}^{-n}C_k(-2ia)^{-(n+k)}\tilde\a^{t\to -ia}_{k-n}
\eea
If there is no other operator which maps to the inside of the contour, the requirement to obtain a nonzero result imposed by acting $\a^{t\to ia}_{k-n}$ on the vacuum defined at $t=ia$, 
\be
\a^{t\to ia}_{k-n}|0\rangle\neq 0
\ee
is that
\be
k - n < 0~~\implies~~ k < n
\ee
This gives
\bea\label{boson prime initial} 
\a'^{(1)}_{-n} = \sum_{k=0}^{n-1}{}^{-n}C_k\,(2ia)^{-(n+k)}\tilde\a^{t\to ia}_{k-n}
\eea
and similarly for copy 2
\bea\label{boson prime initial 2} 
\a'^{(2)}_{-n} = \sum_{k=0}^{n-1}{}^{-n}C_k\,(-2ia)^{-(n+k)}\tilde\a^{t\to -ia}_{k-n}
\eea
These modes obey the commutation relations (\ref{init comm t}).
Next we expand the mode sitting at $t=ia$ to $t=\infty$. To do so we first write the mode on the RHS of (\ref{boson prime initial})
\bea\label{mode at ia}
\tilde\a^{t\to ia}_{k-n} = {1\over2\pi}\oint_{C_{ia}}dt(t-ia)^{k-n}\partial X(t)
\eea 
Let's expand the integrand at $t=\infty$
\bea\label{ia to infty}
(t-ia)^{k-n}&=&t^{k-n}(1-iat^{-1})^{k-n}\cr
&=&\sum_{k'\geq 0}{}^{k-n}C_{k'}(-ia)^{k'}t^{k-n-k'}
\eea
Inserting (\ref{ia to infty}) into (\ref{mode at ia}) and using (\ref{boson t}) yields
\bea
\tilde\a^{t\to ia}_{k-n} = \sum_{k'\geq 0}{}^{k-n}C_{k'}(-ia)^{k'}\tilde{\a}^{t\to\infty}_{k-n-k'}
\eea
Inserting this into (\ref{boson prime initial}) gives
\bea \label{prime from ini to infty}
\a'^{(1)}_{-n} = \sum_{k=0}^{n-1}\sum_{k'\geq 0}{}^{-n}C_k\,{}^{k-n}C_{k'}\,(2ia)^{-(n+k)}(-ia)^{k'}\tilde{\a}^{t\to\infty}_{k-n-k'}
\eea
Again, if there is no other operator which maps to the inside of the contour, to obtain a nonzero result we require that 
\bea
\tilde{\a}^{t\to\infty}_{k-n-k'}|0\rangle_t\neq 0
\eea
which implies that 
\bea 
q = k-n-k' < 0\implies k'=k-n-q
\eea 
Since $k'\geq0$ we find that
\bea  
k-n-q \geq 0\implies k\geq n+q
\eea
For the sum over $k$ we have the following ranges
\bea 
n+q \leq 0&:& \qquad 0\leq k \leq n-1\cr
n+q \geq 0&:& \qquad n+q\leq k \leq n-1
\eea
The mode in (\ref{prime from ini to infty}) can thus be written as
\bea 
\label{prime from ini to infty 2}
\a'^{(1)}_{-n} = \sum_{q\leq-1}\bigg(\sum_{k=\max[0,n+q]}^{n-1}{}^{-n}C_k{}^{k-n}C_{k-n-q}(2ia)^{-(n+k)}(-ia)^{k-n-q}\bigg)\tilde{\a}^{t\to\infty}_{q}
\eea
Performing the sum in parenthesis gives 
\bea \label{prime from ini to infty 3}
\a'^{(1)}_{-n} =\sum_{q\leq-1}{(-1)^ni^{-q}a^{-2n-q}\Gamma(-{q\over2})\over2\Gamma(n)\Gamma(1-(n+{q\over2}))}\tilde{\a}^{t\to\infty}_{q}
\eea 
Inserting (\ref{prime infty}) and (\ref{prime from ini to infty 3}) into (\ref{fi}) for copy $1$ gives
\bea
f_1[-n,-p] &=&  {1\over2p}\,{}_t\langle 0|\a'_{p}\a'^{(1)}_{-n}|0\rangle_t\cr 
&=&   {1\over2p}\sum_{q\leq-1}{(-1)^ni^{-q}a^{-2n-q}\Gamma(-{q\over2})\over2\Gamma(n)\Gamma(1-(n+{q\over2}))}\sum_{j\geq 0}{}^{p}C_jz_0^{j}\big[\tilde\a^{t\to\infty}_{2p-2j},\tilde{\a}^{t\to\infty}_{q}\big]
\cr
&=& {1\over2p}\sum_{q\leq-1}{(-1)^ni^{-q}a^{-2n-q}\Gamma(-{q\over2})\over2\Gamma(n)\Gamma(1-(n+{q\over2}))}\sum_{j\geq 0}{}^{p}C_jz_0^{j}(2p-2j)\d_{2p-2j+q,0}
\eea
The delta function constraint gives 
\bea\label{mode constraints} 
0&=& 2p-2j+q~~\implies~~j= p+{q\over2} \cr
j&\geq& 0\implies  p+{q\over2}\geq 0\implies q\geq -2p
\eea
Making these substitutions give
\be 
f_1[-n,-p]=  {a^{2p-2n}\over2p}\sum_{q=-2p}^{-1}{(-1)^ni^{-q}\Gamma(-{q\over2})(-{q\over2})\over\Gamma(n)\Gamma(1-(n+{q\over2}))}\,{}^{p}C_{p+{q\over2}}
\ee
Since the index $j$ must be an integer (\ref{mode constraints}) indicates that if $p$ is an integer then $q$ is required to be an even integer and if $p$ is half integer then $q$ is required to be an odd integer. We first consider the case where $p$ is an integer and $q$ is an even integer.
\be
p=p',\qquad q=2q',\qquad p',q'\in\mathbb{Z}
\ee
Our result becomes
\be\label{f int}
f_1[-n,-p']=  {a^{2p'-2n}\over2p'}\sum_{q'=-p'}^{-1}{(-1)^ni^{-2q'}\Gamma(-q'+1)\over\Gamma(n)\Gamma(1-(n+q'))}~{}^{p'}C_{p'+q'}
={1\over2} \delta_{n,p'}
\ee
Now consider when $p$ is half integer which also requires $q$ to be odd 
\be
p=p' - {1\over2},\qquad q = 2q'+1
\ee
Inserting this into the above and simplifying gives
\bea
f_1[-n,-p'+\frac{1}{2}] &=&   -ia^{2p'-2n-1}\sum_{q'=-p'}^{-1}{(-1)^{n+q'}\Gamma(-{1\over2}+p')\over2\Gamma(n)\Gamma({1\over2}-(n+q'))\Gamma(1+q'+p')}
\cr 
&=&{ia^{2p'-2n-1}\Gamma({1\over2}+n)\Gamma(-{1\over2}+p')\over\pi(-1 - 2n +2p')\Gamma(n)\Gamma(p')}
\eea
which is
\bea\label{c f1}
f_1[-n,-p]
&=&{ia^{2p-2n}\Gamma({1\over2}+n)\Gamma(p)\over\pi(2p - 2n)\Gamma(n)\Gamma(p+\frac{1}{2})}
\eea

For copy $2$ we simply take $a\to-a$. For integer $p$ the result is the same as (\ref{f int})
\bea
f_2[-n,-p]&=&{1\over2}\delta_{n,p}
\eea
for half integer $p$ the result changes by a minus sign
\bea\label{c f2}
f_2[-n,-p]&=&-{ia^{2p-2n}\Gamma({1\over2}+n)\Gamma(p)\over\pi(2p - 2n)\Gamma(n)\Gamma(p+\frac{1}{2})}
\eea

\subsection{Contraction}

Here we compute the contraction terms $C^{ij}$ using the covering map. We start with two modes on copy 1 in the initial state and compute the following amplitude
\be
C^{11}[n_1,n_2]= \langle0^2|\s(z_0)\a^{(1)}_{-n_1}\a^{(1)}_{-n_2}|0\rangle^{(1)}|0\rangle^{(2)}
\ee
Mapping to the $t$-plane using (\ref{cover map}) gives
\be
C^{11}[n_1,n_2]={}_t\langle0|\a'^{(1)}_{-n_1}\a'^{(1)}_{-n_2}|0\rangle_t
\ee
where the primed modes are defined in (\ref{boson prime initial}) and (\ref{boson prime initial 2}).  The contour of the mode labeled by $n_1$ is mapped to the outside of the contour of the mode labeled by $n_2$. To obtain a contraction, the inside contour should give negative modes and the outside contour should give positive modes. 
Our expression becomes
\be
C^{11}[n_1,n_2]=\sum_{j> n_1}\sum_{k=0}^{n_2-1}{}^{-n_1}C_j{}^{-n_2}C_k(2ia)^{-(n_1+j)}(2ia)^{-(n_2+k)}{}_t\langle 0|\tilde\a^{t\to ia}_{j-n_1}\,\tilde\a^{t\to ia}_{k-n_2}| 0\rangle_t
\ee
Using commutation relations (\ref{init comm t}) we find that
\be
C^{11}[n_1,n_2]=-\sum_{k=0}^{n_2-1}{}^{-n_1}C_{n_1+n_2-k}{}^{-n_2}C_k(2ia)^{-(2n_1 + n_2 - k)}(2ia)^{-(n_2+k)}(k-n_2)
\ee
Performing the sum gives
\bea
C^{11}[n_1,n_2] ={a^{-2(n_1+n_2)}\over2(n_1+n_2)\pi}{\Gamma({1\over2} + n_1)\Gamma({1\over2} + n_2)\over\Gamma(n_1)\Gamma(n_2)} 
\eea
Notice that this expression is symmetric between $n_1$ and $n_2$.
By switching $a\to-a$, we find
\bea
C^{22}[n_1,n_2] = C^{11}[n_1,n_2]
\eea

Let's compute the contraction $C^{12}$. We start with
\bea\label{A-60}
C^{12}[n_1,n_2]\!\!\!&=&\!\!\!{}_t\langle0|\a'^{(1)}_{-n_1}\a'^{(2)}_{-n_2}|0\rangle_t\nn
\!\!\!&=&\!\!\!\sum^{n_1-1}_{j= 0}\sum_{k=0}^{n_2-1}{}^{-n_1}C_j{}^{-n_2}C_k(2ia)^{-(n_1+j)}(2ia)^{-(n_2+k)}
{}_t\langle 0|\tilde\a^{t\to ia}_{j-n_1}\,\tilde\a^{t\to -ia}_{k-n_2}|0\rangle_t
\eea
On the $t$-plane, the contour of copy 1 is located at $t=ia$ and the contour of copy 2 is located at $t=-ia$. Both contours should be left with negative modes.
To compute the expectation value, we expand the contour of the mode located at $t=ia$ around the point $t=-ia$. To do this we start with a mode defined at $t=ia$
\be\label{A-62}
\tilde\alpha^{t\to ia}_{m} = \frac{1}{2\pi}\oint_{C_{ia}} dt (t-ia)^m\partial X(t)
\ee
We expand the integrand in the following way
\be
(t-ia)^m = (-2ia+t+ia)^m = \sum_{p\geq 0} {}^{m}C_{p}(-2ia)^{m-p}(t+ia)^{p}
\ee
Inserting this into (\ref{A-62}) gives
\be
\tilde\alpha^{t\to ia}_{m} = (-1) \sum_{p\geq 0} {}^{m}C_{p}(-2ia)^{m-p}\tilde\alpha^{t\to -ia}_{p}
\ee
where we have used the modes defined around $t=-ia$ in (\ref{nat t init}).
Using commutation relations (\ref{init comm t}), the expectation value of a mode at $t=ia$ and a mode at $t=-ia$ becomes
\be
{}_t\langle 0|\tilde\a^{t\to ia}_{m}\,\tilde\a^{t\to -ia}_{n}|0\rangle_t =
(-1) \sum_{p\geq 0} {}^{m}C_{p}(-2ia)^{m-p} p \delta_{p+n,0}
=  {}^{m}C_{-n}(-2ia)^{m+n} n 
\ee
Using this expression in (\ref{A-60}) gives
\bea
C^{12}[n_1,n_2]
=(2a)^{-2(n_1+n_2)}\sum^{n_1-1}_{j= 0}\sum_{k=0}^{n_2-1}
{}^{-n_1}C_j{}^{-n_2}C_k {}^{j-n_1}C_{n_2-k}
(-1)^{j+k} (k-n_2)
\eea
Performing the sums give
\bea
C^{12}[n_1,n_2] =-{a^{-2(n_1+n_2)}\over2(n_1+n_2)\pi}{\Gamma({1\over2} + n_1)\Gamma({1\over2} + n_2)\over\Gamma(n_1)\Gamma(n_2)} 
\eea
Notice that this expression is symmetric between $n_1$ and $n_2$.
By switching $a\to-a$, we find
\bea
C^{12}[n_1,n_2] = C^{21}[n_1,n_2]
\eea
Therefore, all the contractions can be written as
\bea
C^{ij}[n_1,n_2] =(-1)^{i+j}{a^{-2(n_1+n_2)}\over2(n_1+n_2)\pi}{\Gamma({1\over2} + n_1)\Gamma({1\over2} + n_2)\over\Gamma(n_1)\Gamma(n_2)} 
\eea
where $i,j=1,2$ are copy labels.

\section{Propagation: higher modes}\label{App f}

In this appendix we will derive the propagation $f_i[-n,-p]$ with $n>1$ from $f_i[-1,-p]$ by applying $L_{-1}$ repeatedly.
For copy 1 we begin with following expression
\bea
\s(z_0)\a^{(1)}_{-n}|0\rangle^{(1)}|0\rangle^{(2)}={1\over\Gamma(n)}\s(z_0)(L_{-1})^{n-1}\a^{(1)}_{-1}|0\rangle^{(1)}|0\rangle^{(2)}
\eea
Using relation (\ref{L-1 s}) we obtain 
\bea
\s(z_0)\a^{(1)}_{-n}|0\rangle^{(1)}|0\rangle^{(2)}={1\over\Gamma(n)}(L_{-1} - \partial)^{n-1}\s(z_0)\a^{(1)}_{-1}|0\rangle^{(1)}|0\rangle^{(2)}
\eea
Using (\ref{one}) we pass the bosonic mode through the twist on both sides and only keep the terms with a single mode. This gives 
\bea\label{relation} 
\sum_{p> 0}f_1[-n,-p]\a_{-p}|0^2\rangle&=&{1\over\Gamma(n)}(L_{-1} - \partial)^{n-1}\sum_{p'> 0}f_1[-1,-p']\a_{-p'}|0^2\rangle
\eea
Let's compute the RHS. We look at the term
\bea 
(L_{-1} - \partial)^{n-1}f_1[-1,-p']\a_{-p'}
= \sum_{k\geq0}(L_{-1})^k\a_{-p'}{}^{n-1}C_k(-1)^{n-k-1}\partial^{\,n-k-1}f_1[-1,-p']
\eea 
We only keep terms where $L_{-1}$ acts on $\a_{-p'}$ since this leaves us with just one mode. 
Only keeping terms which will leave us with one mode and using the expression in (\ref{f1pc1}) and (\ref{f2pc1}) we have 
\bea
&&\!\!(L_{-1} - \partial )^{n-1}f_1[-1,-p']\a_{-p'}\cr
&&\!\!=f_1[-1,-p']_{z_0=1}\sum_{k=0}^{n-1}(L_{-1})^k\a_{-p'}\,{}^{n-1}C_k(-1)^{n-k-1}\partial^{\,n-k-1}z_0^{p'-1}\cr
&&\!\!=f_1[-1,-p']_{z_0=1}\sum_{k=0}^{n-1}{\Gamma(p'+k)\over\Gamma(p')}\,{}^{n-1}C_k(-1)^{n-k-1}{\Gamma(p')\over\Gamma(p'-(n-k-1))}z_0^{p'+k-n}\a_{-(p'+k)} + \ldots
\cr
&&\!\!=\sum_{k=0}^{n-1}f_1[-1,-p']_{z_0=1}{\Gamma(p'+k)\over\Gamma(p'-(n-k-1))}{}^{n-1}C_k(-1)^{n-k-1}z_0^{p'+k-n}\a_{-(p'+k)}+\ldots
\eea
Inserting this expression back into (\ref{relation}) we find
\bea
\sum_{p> 0}f_1[-n,-p]\a_{-p}|0^2\rangle\!\!&=&\!\!{1\over\Gamma(n)}\sum_{k=0}^{n-1}\sum_{p'>0}f_1[-1,-p']_{z_0=1}{\Gamma(p'+k)\over\Gamma(p'-(n-k-1))}{}^{n-1}C_k(-1)^{n-k-1}\cr 
&&z_0^{p'+k-n}\a_{-(p'+k)}|0^2\rangle+\ldots
\eea
In order to compare the terms on the left and right hand sides we take $p'+k = p$ which allows us to compare the following terms 
\bea 
f_1[-n,-p] \!&=&\!Cz_0^{p-n}{\Gamma(p)\over\Gamma(n)\Gamma(p-(n-1))}\sum_{k=0}^{n-1}\frac{\Gamma(p-k-1)}{\Gamma(p-k+1/2)}{}^{n-1}C_k(-1)^{n-k-1}\cr\cr
\!&=&\!Cz_0^{p-n}2(-1)^n\sqrt\pi{\csc(\pi p)\Gamma(p)\Gamma({1\over2}+n)\over\Gamma(n)\Gamma(1-n+p)\Gamma(1+n-p)\Gamma({1\over2} + p)}
\eea
Since $p$ is a half integer, using the expression for $C$, (\ref{C}), we obtain 
\bea\label{fi} 
f_1[-n,-p]&=&{iz_0^{p-n}\Gamma(p)\Gamma({1\over2}+n)\over\pi(2p-2n)\Gamma(p+\frac{1}{2})\Gamma(n)}
\eea
where $n$ is an integer and $p$ is a half integer.
To obtain $f_2[-n,-p]$ for an excitation on copy 2, we change the location of the twist by $z_0\to z_0 e^{2\pi i}$. It interchanges copy 1 and copy 2. Since $p-n$ is a half integer, we have $z_0^{p-1} \to -z_0^{p-1}$, which gives
\bea\label{fi} 
f_2[-n,-p]&=&-f_1[-n,-p]
\eea
Notice that if we take $z_0=a^2$ this result agrees with the result from the covering map method (\ref{c f1}) and (\ref{c f2}).

\section{Contraction: higher modes}\label{Cij}

Here we derive $C^{ij}[n_1,n_2]$.
We do this by looking at the following relation
\be
\s(z_0)\a^{(i)}_{-n_1}\a^{(j)}_{-n_2}|0\rangle^{(1)}|0\rangle^{(2)}={1\over\Gamma(n_2)}\s(z_0)\a^{(i)}_{-n_1}(L_{-1})^{n_2-1}\a^{(j)}_{-1}|0\rangle^{(1)}|0\rangle^{(2)}
\ee
To compute the RHS, notice that
\be
\a^{(i)}_{-n_1} L_{-1} =  (L_{-1}-L_{-1}\circ) \a^{(i)}_{-n_1}
\ee
where 
\be
L_{-1}\circ O_{-n} = [L_{-1}, O_{-n}]
\ee
Thus, we have
\bea
\a^{(i)}_{-n_1} L^{n_2-1}_{-1} &=&  (L_{-1}-L_{-1}\circ)^{n_2-1} \a^{(i)}_{-n_1}\nn
&=&\sum_{k=0}^{n_2-1}{}^{n_2-1}C_k(-1)^{n_2-1-k}(L_{-1})^k(L_{-1}\circ)^{n_2-1-k}\a^{(i)}_{-n_1}
\eea
where we have used the fact that $L_{-1}\circ L_{-1}=0$.
Therefore, we obtain
\bea 
\s(z_0)\a^{(i)}_{-n_1}\a^{(j)}_{-n_2}|0\rangle^{(1)}|0\rangle^{(2)}
&=&{1\over\Gamma(n_2)}\sum_{k=0}^{n_2-1}{}^{n_2-1}C_k{\Gamma(n_1+n_2-1-k)\over\Gamma(n_1)}(-1)^{n_2-1-k}\cr
&&\s(z_0)(L_{-1})^k\a^{(i)}_{-(n_1+n_2-1-k)}\a^{(j)}_{-1}|0\rangle^{(1)}|0\rangle^{(2)}
\eea 
Using the relation (\ref{L-1 s}) we have
\bea
\s(z_0)\a^{(i)}_{-n_1}\a^{(j)}_{-n_2}|0\rangle^{(1)}|0\rangle^{(2)}&=&{1\over\Gamma(n_1)\Gamma(n_2)}\sum_{k=0}^{n_2-1}{}^{n_2-1}C_k\Gamma(n_1+n_2-1-k)(-1)^{n_2-1-k}\cr
&&(L_{-1}-\partial )^k\s(z_0)\a^{(i)}_{-(n_1+n_2-1-k)}\a^{(j)}_{-1}|0\rangle^{(1)}|0\rangle^{(2)}
\eea
Keeping the terms which contain no modes on both the LHS and RHS we find 
\bea\label{C11n1n2}
C^{ij}[n_1,n_2] &=& {1\over\Gamma(n_1)\Gamma(n_2)}\sum_{k=0}^{n_2-1}{}^{n_2-1}C_k\Gamma(n_1+n_2-1-k)(-1)^{n_2-1-k}\cr
&&(-\partial)^k C^{ij}[n_1+n_2-1-k,1]
\eea
Notice that $C^{ij}[1,n] = C^{ji}[n,1]$ since the contraction is between two bosonic modes whose order can be changed. Using the expression in (\ref{C111n})
we find that (\ref{C11n1n2}) becomes
\bea \label{C11n1n2p}
C^{ij}[n_1,n_2] = {(-1)^{i+j}(-1)^{n_2-1}\over4\sqrt\pi\Gamma(n_1)\Gamma(n_2)}\sum_{k=0}^{n_2-1}{}^{n_2-1}C_k{\Gamma(-{1\over2}+n_1+n_2-k)\over(n_1+n_2-k)}\partial^k z_0^{-(n_1+n_2-k)}
\eea
which gives
\be
C^{ij}[n_1,n_2]
= (-1)^{i+j}{z_0^{-(n_1+n_2)}\over2(n_1+n_2)\pi}{\Gamma({1\over2}+n_1)\Gamma({1\over2}+n_2)\over\Gamma(n_1)\Gamma(n_2)}
\ee
where $i,j=1,2$ are copy labels.

\bibliographystyle{JHEP}
\bibliography{bibliography.bib}

\providecommand{\href}[2]{#2}\begingroup\raggedright\begin{thebibliography}{10}

\bibitem{Maldacena:1997re}
J.M.~Maldacena, \emph{{The Large N limit of superconformal field theories and
  supergravity}}, \href{https://doi.org/10.1023/A:1026654312961}{\emph{Adv.
  Theor. Math. Phys.} {\bfseries 2} (1998) 231}
  [\href{https://arxiv.org/abs/hep-th/9711200}{{\ttfamily hep-th/9711200}}].

\bibitem{Strominger:1996sh}
A.~Strominger and C.~Vafa, \emph{{Microscopic origin of the Bekenstein-Hawking
  entropy}}, \href{https://doi.org/10.1016/0370-2693(96)00345-0}{\emph{Phys.
  Lett. B} {\bfseries 379} (1996) 99}
  [\href{https://arxiv.org/abs/hep-th/9601029}{{\ttfamily hep-th/9601029}}].

\bibitem{Maldacena:1999bp}
J.M.~Maldacena, G.W.~Moore and A.~Strominger, \emph{{Counting BPS black holes
  in toroidal Type II string theory}},
  \href{https://arxiv.org/abs/hep-th/9903163}{{\ttfamily hep-th/9903163}}.

\bibitem{Seiberg:1999xz}
N.~Seiberg and E.~Witten, \emph{{The D1 / D5 system and singular CFT}},
  \href{https://doi.org/10.1088/1126-6708/1999/04/017}{\emph{JHEP} {\bfseries
  04} (1999) 017} [\href{https://arxiv.org/abs/hep-th/9903224}{{\ttfamily
  hep-th/9903224}}].

\bibitem{Dijkgraaf:1998gf}
R.~Dijkgraaf, \emph{{Instanton strings and hyperKahler geometry}},
  \href{https://doi.org/10.1016/S0550-3213(98)00869-4}{\emph{Nucl. Phys. B}
  {\bfseries 543} (1999) 545}
  [\href{https://arxiv.org/abs/hep-th/9810210}{{\ttfamily hep-th/9810210}}].

\bibitem{Larsen:1999uk}
F.~Larsen and E.J.~Martinec, \emph{{U(1) charges and moduli in the D1 - D5
  system}}, \href{https://doi.org/10.1088/1126-6708/1999/06/019}{\emph{JHEP}
  {\bfseries 06} (1999) 019}
  [\href{https://arxiv.org/abs/hep-th/9905064}{{\ttfamily hep-th/9905064}}].

\bibitem{Jevicki:1998bm}
A.~Jevicki, M.~Mihailescu and S.~Ramgoolam, \emph{{Gravity from CFT on S**N(X):
  Symmetries and interactions}},
  \href{https://doi.org/10.1016/S0550-3213(00)00147-4}{\emph{Nucl. Phys. B}
  {\bfseries 577} (2000) 47}
  [\href{https://arxiv.org/abs/hep-th/9907144}{{\ttfamily hep-th/9907144}}].

\bibitem{deBoer:1998kjm}
J.~de~Boer, \emph{{Six-dimensional supergravity on S**3 x AdS(3) and 2-D
  conformal field theory}},
  \href{https://doi.org/10.1016/S0550-3213(99)00160-1}{\emph{Nucl. Phys. B}
  {\bfseries 548} (1999) 139}
  [\href{https://arxiv.org/abs/hep-th/9806104}{{\ttfamily hep-th/9806104}}].

\bibitem{Lunin:2000yv}
O.~Lunin and S.D.~Mathur, \emph{{Correlation functions for M**N / S(N)
  orbifolds}}, \href{https://doi.org/10.1007/s002200100431}{\emph{Commun. Math.
  Phys.} {\bfseries 219} (2001) 399}
  [\href{https://arxiv.org/abs/hep-th/0006196}{{\ttfamily hep-th/0006196}}].

\bibitem{Lunin:2001pw}
O.~Lunin and S.D.~Mathur, \emph{{Three point functions for M(N) / S(N)
  orbifolds with N=4 supersymmetry}},
  \href{https://doi.org/10.1007/s002200200638}{\emph{Commun. Math. Phys.}
  {\bfseries 227} (2002) 385}
  [\href{https://arxiv.org/abs/hep-th/0103169}{{\ttfamily hep-th/0103169}}].

\bibitem{Pakman:2009zz}
A.~Pakman, L.~Rastelli and S.S.~Razamat, \emph{{Diagrams for Symmetric Product
  Orbifolds}}, \href{https://doi.org/10.1088/1126-6708/2009/10/034}{\emph{JHEP}
  {\bfseries 10} (2009) 034} [\href{https://arxiv.org/abs/0905.3448}{{\ttfamily
  0905.3448}}].

\bibitem{Pakman:2009ab}
A.~Pakman, L.~Rastelli and S.S.~Razamat, \emph{{Extremal Correlators and
  Hurwitz Numbers in Symmetric Product Orbifolds}},
  \href{https://doi.org/10.1103/PhysRevD.80.086009}{\emph{Phys. Rev. D}
  {\bfseries 80} (2009) 086009}
  [\href{https://arxiv.org/abs/0905.3451}{{\ttfamily 0905.3451}}].

\bibitem{Avery:2010er}
S.G.~Avery, B.D.~Chowdhury and S.D.~Mathur, \emph{{Deforming the D1D5 CFT away
  from the orbifold point}},
  \href{https://doi.org/10.1007/JHEP06(2010)031}{\emph{JHEP} {\bfseries 06}
  (2010) 031} [\href{https://arxiv.org/abs/1002.3132}{{\ttfamily 1002.3132}}].

\bibitem{Avery:2010hs}
S.G.~Avery, B.D.~Chowdhury and S.D.~Mathur, \emph{{Excitations in the deformed
  D1D5 CFT}}, \href{https://doi.org/10.1007/JHEP06(2010)032}{\emph{JHEP}
  {\bfseries 06} (2010) 032} [\href{https://arxiv.org/abs/1003.2746}{{\ttfamily
  1003.2746}}].

\bibitem{Avery:2010qw}
S.G.~Avery, \emph{{Using the D1D5 CFT to Understand Black Holes}},  other
  thesis, 12, 2010, [\href{https://arxiv.org/abs/1012.0072}{{\ttfamily
  1012.0072}}].

\bibitem{Carson:2014yxa}
Z.~Carson, S.~Hampton, S.D.~Mathur and D.~Turton, \emph{{Effect of the twist
  operator in the D1D5 CFT}},
  \href{https://doi.org/10.1007/JHEP08(2014)064}{\emph{JHEP} {\bfseries 08}
  (2014) 064} [\href{https://arxiv.org/abs/1405.0259}{{\ttfamily 1405.0259}}].

\bibitem{Carson:2014ena}
Z.~Carson, S.~Hampton, S.D.~Mathur and D.~Turton, \emph{{Effect of the
  deformation operator in the D1D5 CFT}},
  \href{https://doi.org/10.1007/JHEP01(2015)071}{\emph{JHEP} {\bfseries 01}
  (2015) 071} [\href{https://arxiv.org/abs/1410.4543}{{\ttfamily 1410.4543}}].

\bibitem{Carson:2014xwa}
Z.~Carson, S.D.~Mathur and D.~Turton, \emph{{Bogoliubov coefficients for the
  twist operator in the D1D5 CFT}},
  \href{https://doi.org/10.1016/j.nuclphysb.2014.10.018}{\emph{Nucl. Phys. B}
  {\bfseries 889} (2014) 443}
  [\href{https://arxiv.org/abs/1406.6977}{{\ttfamily 1406.6977}}].

\bibitem{Carson:2017byr}
Z.~Carson, I.T.~Jardine and A.W.~Peet, \emph{{Component twist method for higher
  twists in D1-D5 CFT}},
  \href{https://doi.org/10.1103/PhysRevD.96.026006}{\emph{Phys. Rev. D}
  {\bfseries 96} (2017) 026006}
  [\href{https://arxiv.org/abs/1704.03401}{{\ttfamily 1704.03401}}].

\bibitem{Burrington:2014yia}
B.A.~Burrington, S.D.~Mathur, A.W.~Peet and I.G.~Zadeh, \emph{{Analyzing the
  squeezed state generated by a twist deformation}},
  \href{https://doi.org/10.1103/PhysRevD.91.124072}{\emph{Phys. Rev. D}
  {\bfseries 91} (2015) 124072}
  [\href{https://arxiv.org/abs/1410.5790}{{\ttfamily 1410.5790}}].

\bibitem{Dixon:1986qv}
L.J.~Dixon, D.~Friedan, E.J.~Martinec and S.H.~Shenker, \emph{{The Conformal
  Field Theory of Orbifolds}},
  \href{https://doi.org/10.1016/0550-3213(87)90676-6}{\emph{Nucl. Phys. B}
  {\bfseries 282} (1987) 13}.

\bibitem{Arutyunov:1997gt}
G.E.~Arutyunov and S.A.~Frolov, \emph{{Virasoro amplitude from the S**N R**24
  orbifold sigma model}},
  \href{https://doi.org/10.1007/BF02557107}{\emph{Theor. Math. Phys.}
  {\bfseries 114} (1998) 43}
  [\href{https://arxiv.org/abs/hep-th/9708129}{{\ttfamily hep-th/9708129}}].

\bibitem{Arutyunov:1997gi}
G.E.~Arutyunov and S.A.~Frolov, \emph{{Four graviton scattering amplitude from
  S**N R**8 supersymmetric orbifold sigma model}},
  \href{https://doi.org/10.1016/S0550-3213(98)00326-5}{\emph{Nucl. Phys. B}
  {\bfseries 524} (1998) 159}
  [\href{https://arxiv.org/abs/hep-th/9712061}{{\ttfamily hep-th/9712061}}].

\bibitem{Dei:2019iym}
A.~Dei and L.~Eberhardt, \emph{{Correlators of the symmetric product
  orbifold}}, \href{https://doi.org/10.1007/JHEP01(2020)108}{\emph{JHEP}
  {\bfseries 01} (2020) 108}
  [\href{https://arxiv.org/abs/1911.08485}{{\ttfamily 1911.08485}}].

\bibitem{Gava:2002xb}
E.~Gava and K.S.~Narain, \emph{{Proving the PP wave / CFT(2) duality}},
  \href{https://doi.org/10.1088/1126-6708/2002/12/023}{\emph{JHEP} {\bfseries
  12} (2002) 023} [\href{https://arxiv.org/abs/hep-th/0208081}{{\ttfamily
  hep-th/0208081}}].

\bibitem{Gaberdiel:2015uca}
M.R.~Gaberdiel, C.~Peng and I.G.~Zadeh, \emph{{Higgsing the stringy higher spin
  symmetry}}, \href{https://doi.org/10.1007/JHEP10(2015)101}{\emph{JHEP}
  {\bfseries 10} (2015) 101}
  [\href{https://arxiv.org/abs/1506.02045}{{\ttfamily 1506.02045}}].

\bibitem{Burrington:2018upk}
B.A.~Burrington, I.T.~Jardine and A.W.~Peet, \emph{{The OPE of bare twist
  operators in bosonic $S_N$ orbifold CFTs at large $N$}},
  \href{https://doi.org/10.1007/JHEP08(2018)202}{\emph{JHEP} {\bfseries 08}
  (2018) 202} [\href{https://arxiv.org/abs/1804.01562}{{\ttfamily
  1804.01562}}].

\bibitem{Hampton:2018ygz}
S.~Hampton, S.D.~Mathur and I.G.~Zadeh, \emph{{Lifting of D1-D5-P states}},
  \href{https://doi.org/10.1007/JHEP01(2019)075}{\emph{JHEP} {\bfseries 01}
  (2019) 075} [\href{https://arxiv.org/abs/1804.10097}{{\ttfamily
  1804.10097}}].

\bibitem{DeBeer:2019oxm}
T.~De~Beer, B.A.~Burrington, I.T.~Jardine and A.W.~Peet, \emph{{The large $N$
  limit of OPEs in symmetric orbifold CFTs with $\mathcal{N}=(4,4)$
  supersymmetry}},
  \href{https://doi.org/10.1007/s13130-019-11019-2}{\emph{JHEP} {\bfseries 08}
  (2019) 015} [\href{https://arxiv.org/abs/1904.07816}{{\ttfamily
  1904.07816}}].

\bibitem{Guo:2019pzk}
B.~Guo and S.D.~Mathur, \emph{{Lifting of states in 2-dimensional $N = 4$
  supersymmetric CFTs}},
  \href{https://doi.org/10.1007/JHEP10(2019)155}{\emph{JHEP} {\bfseries 10}
  (2019) 155} [\href{https://arxiv.org/abs/1905.11923}{{\ttfamily
  1905.11923}}].

\bibitem{Keller:2019yrr}
C.A.~Keller and I.G.~Zadeh, \emph{{Conformal Perturbation Theory for Twisted
  Fields}}, \href{https://doi.org/10.1088/1751-8121/ab6b91}{\emph{J. Phys. A}
  {\bfseries 53} (2020) 095401}
  [\href{https://arxiv.org/abs/1907.08207}{{\ttfamily 1907.08207}}].

\bibitem{Hampton:2019hya}
S.~Hampton, \emph{{Understanding Black Hole Formation in String Theory}}, Ph.D.
  thesis, Ohio State U., 2019.
\newblock \href{https://arxiv.org/abs/1909.09310}{{\ttfamily 1909.09310}}.

\bibitem{Hampton:2019csz}
S.~Hampton and S.D.~Mathur, \emph{{Thermalization in the D1D5 CFT}},
  \href{https://doi.org/10.1007/JHEP06(2020)004}{\emph{JHEP} {\bfseries 06}
  (2020) 004} [\href{https://arxiv.org/abs/1910.01690}{{\ttfamily
  1910.01690}}].

\bibitem{Guo:2019ady}
B.~Guo and S.D.~Mathur, \emph{{Lifting of level-1 states in the D1D5 CFT}},
  \href{https://doi.org/10.1007/JHEP03(2020)028}{\emph{JHEP} {\bfseries 03}
  (2020) 028} [\href{https://arxiv.org/abs/1912.05567}{{\ttfamily
  1912.05567}}].

\bibitem{Guo:2020gxm}
B.~Guo and S.D.~Mathur, \emph{{Lifting at higher levels in the D1D5 CFT}},
  \href{https://doi.org/10.1007/JHEP11(2020)145}{\emph{JHEP} {\bfseries 11}
  (2020) 145} [\href{https://arxiv.org/abs/2008.01274}{{\ttfamily
  2008.01274}}].

\bibitem{Lima:2020boh}
A.A.~Lima, G.M.~Sotkov and M.~Stanishkov, \emph{{Microstate Renormalization in
  Deformed D1-D5 SCFT}},
  \href{https://doi.org/10.1016/j.physletb.2020.135630}{\emph{Phys. Lett. B}
  {\bfseries 808} (2020) 135630}
  [\href{https://arxiv.org/abs/2005.06702}{{\ttfamily 2005.06702}}].

\bibitem{Lima:2020nnx}
A.A.~Lima, G.M.~Sotkov and M.~Stanishkov, \emph{{Correlation functions of
  composite Ramond fields in deformed D1-D5 orbifold SCFT$_2$}},
  \href{https://doi.org/10.1103/PhysRevD.102.106004}{\emph{Phys. Rev. D}
  {\bfseries 102} (2020) 106004}
  [\href{https://arxiv.org/abs/2006.16303}{{\ttfamily 2006.16303}}].

\bibitem{Lima:2020urq}
A.A.~Lima, G.M.~Sotkov and M.~Stanishkov, \emph{{Dynamics of R-neutral Ramond
  fields in the D1-D5 SCFT}},
  \href{https://arxiv.org/abs/2012.08021}{{\ttfamily 2012.08021}}.

\bibitem{Lima:2020kek}
A.A.~Lima, G.M.~Sotkov and M.~Stanishkov, \emph{{Renormalization of twisted
  Ramond fields in D1-D5 SCFT$_{2}$}},
  \href{https://doi.org/10.1007/JHEP03(2021)202}{\emph{JHEP} {\bfseries 03}
  (2021) 202} [\href{https://arxiv.org/abs/2010.00172}{{\ttfamily
  2010.00172}}].

\bibitem{Lima:2021wrz}
A.A.~Lima, G.M.~Sotkov and M.~Stanishkov, \emph{{On the Dynamics of Protected
  Ramond Ground States in the D1-D5 CFT}},
  \href{https://arxiv.org/abs/2103.04459}{{\ttfamily 2103.04459}}.

\bibitem{Benjamin:2021zkn}
N.~Benjamin, C.A.~Keller and I.G.~Zadeh, \emph{{Lifting 1/4-BPS states in
  AdS$_{3}$$\times$ S$^{3}$ $\times$ T$^{4}$}},
  \href{https://doi.org/10.1007/JHEP10(2021)089}{\emph{JHEP} {\bfseries 10}
  (2021) 089} [\href{https://arxiv.org/abs/2107.00655}{{\ttfamily
  2107.00655}}].

\bibitem{AlvesLima:2022elo}
A.~Alves~Lima, G.M.~Sotkov and M.~Stanishkov, \emph{{Four-point functions with
  multi-cycle fields in symmetric orbifolds and the D1-D5 CFT}},
  \href{https://doi.org/10.1007/JHEP05(2022)106}{\emph{JHEP} {\bfseries 05}
  (2022) 106} [\href{https://arxiv.org/abs/2202.12424}{{\ttfamily
  2202.12424}}].

\bibitem{Apolo:2022fya}
L.~Apolo, A.~Belin, S.~Bintanja, A.~Castro and C.A.~Keller, \emph{{Deforming
  Symmetric Product Orbifolds: A tale of moduli and higher spin currents}},
  \href{https://arxiv.org/abs/2204.07590}{{\ttfamily 2204.07590}}.

\bibitem{Carson:2016cjj}
Z.~Carson, S.~Hampton and S.D.~Mathur, \emph{{One-Loop Transition Amplitudes in
  the D1D5 CFT}}, \href{https://doi.org/10.1007/JHEP01(2017)006}{\emph{JHEP}
  {\bfseries 01} (2017) 006}
  [\href{https://arxiv.org/abs/1606.06212}{{\ttfamily 1606.06212}}].

\bibitem{Carson:2016uwf}
Z.~Carson, S.~Hampton and S.D.~Mathur, \emph{{Full action of two deformation
  operators in the D1D5 CFT}},
  \href{https://doi.org/10.1007/JHEP11(2017)096}{\emph{JHEP} {\bfseries 11}
  (2017) 096} [\href{https://arxiv.org/abs/1612.03886}{{\ttfamily
  1612.03886}}].

\bibitem{Carson:2015ohj}
Z.~Carson, S.~Hampton and S.D.~Mathur, \emph{{Second order effect of twist
  deformations in the D1D5 CFT}},
  \href{https://doi.org/10.1007/JHEP04(2016)115}{\emph{JHEP} {\bfseries 04}
  (2016) 115} [\href{https://arxiv.org/abs/1511.04046}{{\ttfamily
  1511.04046}}].

\bibitem{Guo:2021ybz}
B.~Guo and S.~Hampton, \emph{{A freely falling graviton in the D1D5 CFT}},
  \href{https://arxiv.org/abs/2107.11883}{{\ttfamily 2107.11883}}.

\bibitem{Guo:2021gqd}
B.~Guo and S.~Hampton, \emph{{The Dual of a Tidal Force in the D1D5 CFT}},
  \href{https://arxiv.org/abs/2108.00068}{{\ttfamily 2108.00068}}.

\end{thebibliography}\endgroup

\end{document}